\documentclass[aps,prb,floatfix,amsmath,amssymb,preprint,eqsecnum,nofootinbib]{revtex4}
\usepackage{graphicx}
\usepackage{dcolumn}
\usepackage{bm}
\usepackage{amsmath, amsfonts, amssymb}
\usepackage{pstricks}
\usepackage{amsxtra}
\usepackage{amsthm}

\newcommand{\eqnref}[1]{Eq.~(\ref{#1})} 
\newcommand{\secref}[1]{Section~\ref{#1}}
\newcommand{\hc}{\mathrm{H.c.}}

\newcommand{\beq}{\begin{equation}}
\newcommand{\eeq}{\end{equation}}
\newcommand{\ba}{\begin{array}{ccc}}
\newcommand{\ea}{\end{array}}
\newcommand{\nn}{\nonumber \\}

\def\bea{\begin{eqnarray}}
\def\eea{\end{eqnarray}}
\renewcommand{\l}{\lambda}

\usepackage{setspace} 
\begin{document}
\title{Fermi surfaces and gauge-gravity duality}
\author{Liza Huijse}
\affiliation{Department of Physics, Harvard University, Cambridge MA
02138}
\author{Subir Sachdev}
\affiliation{Department of Physics, Harvard University, Cambridge MA
02138}

\date{\today \\
\vspace{1.6in}}
\begin{abstract}
We give a unified overview of the zero temperature phases of compressible quantum matter: {\em i.e.\/} phases in which the expectation value
of a globally conserved U(1) density, $\mathcal{Q}$, varies smoothly as a function of parameters. Provided the global 
U(1) and translational symmetries are unbroken, such phases are expected to have Fermi surfaces, and the Luttinger theorem 
relates the volumes enclosed by these Fermi surfaces to $\langle \mathcal{Q} \rangle$. We survey models of interacting bosons and/or fermions and/or
gauge fields which realize such phases. Some phases have Fermi surfaces with the singularities of Landau's Fermi liquid theory, while other
Fermi surfaces have non-Fermi liquid singularities.
Compressible phases found in models applicable to condensed matter systems
are argued to also be present in models obtained by applying chemical potentials (and other deformations
allowed by the residual symmetry at non-zero chemical potential)
to the paradigmic supersymmetric gauge theories underlying gauge-gravity duality: the ABJM model in spatial dimension $d=2$,
and the $\mathcal{N}=4$ SYM theory in $d=3$.
\end{abstract}

\maketitle

\section{Introduction}
\label{sec:intro}

There is much recent interest in the topic of {\em compressible quantum matter}. This is motivated partly by the hope
of resolving the puzzle of `strange metal' physics in numerous correlated electron materials. Thinking more
broadly, we may define compressible states of matter as continuum states in spatial dimensions $d>1$ which satisfy the following simple requirements at the absolute zero of temperature ($T=0$):
\begin{itemize}
\item They have a global U(1) symmetry, and an associated conserved density, $\mathcal{Q}$.
\item As we change the value of a `chemical potential' $\mu$ which couples linearly to $\mathcal{Q}$,
the ground state value of $\langle \mathcal{Q} \rangle$
varies smoothly as a function of $\mu$.
\item The global U(1) symmetry and translational symmetry are unbroken in the ground state.
\end{itemize}
Remarkably, there are only a few known states in condensed matter physics which satisfy the above requirements, and 
we will discuss examples of essentially all of them in the present paper. Moreover, all such states have {\em Fermi surfaces\/}, a concept we will define precisely below. The most familiar example of a compressible quantum state is, of course, 
Landau's Fermi liquid, which we will refer to simply as the Fermi liquid (FL). 
It is sometimes assumed that Fermi surfaces occur only in Fermi liquids, but that is not true: Fermi surfaces are more general,
and are present also in other states of matter.

Note that we have not placed any restrictions on the statistics of the microscopic degrees of freedom. The compressible state could be made up of either fermions or bosons, or both. Nevertheless, Fermi surfaces are expected to be present
as long as the global U(1) symmetry is preserved, and the system does not crystallize into a solid by breaking translational symmetry.
The Fermi surfaces could be associated with emergent fermions, which are either composites or fractions of the 
microscopic particles. 

Despite the paucity of known examples of states of compressible matter, such states have proliferated
in recent studies\cite{nernst,sslee0,denef0,hong0,zaanen1,hong1,denef,faulkner,polchinski,gubserrocha,hong2,kiritsis,sean1,sean2,sean3,eric,kachru2,kachru3,kachru4,trivedi,zaanen2} using gauge-gravity duality. Clearly, a proper condensed matter interpretation of these putative states is urgently
needed.\cite{ssffl,mcphys}

\subsection{The Luttinger Theorem}
\label{sec:lutt}

This theorem was originally established for a gas of fermions with weak or moderate interactions. The non-interacting
Fermi gas has a ground state with all states inside the Fermi surface occupied, and so the momentum-space volume enclosed
by the Fermi surface must equal the density of fermions (our momentum space volumes include phase factors of
$(2 \pi)^{-d}$). The Luttinger theorem proves that this Fermi surface volume remains invariant to all
orders in the fermion-fermion interaction. In the early presentations of its proof, it was implicitly assumed that 
a Fermi liquid was under consideration, but the result is actually much more general as we will now discuss.

A more recent discussion of the Luttinger theorem appeared in the works of Powell {\em et al.} \cite{powell,yangs} and
Coleman {\em et al.}, \cite{piers} who applied it to arbitrary interacting systems of fermions, bosons, and gauge fields.
They pointed out the key role played by continuous symmetries and associated global conservation laws, and we will
now review their presentation.

There is a Luttinger theorem for each global U(1) symmetry which is not spontaneously broken, and for simplicity 
let us assume that there is only one:
that associated with the conserved density $\mathcal{Q}$. Express the theory in terms of a complete set of fields 
$\psi_{\ell}$, where $\ell$ is a label identifying the bosons, fermions, or gauge fields. By ``complete set'' we mean
that we include
not only the fundamental canonical fields of the underlying Lagrangian, but also composites or fractions of the fundamental
fields. Composite fields can be introduced via a suitable Hubbard-Stratonovich decoupling of an interaction term, while fractions arise in the slave-particle construction along with emergent gauge fields \cite{tasi}. 
There is no requirement that $\psi_{\ell}$ fields introduced in this manner be canonical.
The complete Lagrangian has a global U(1) symmetry under which
\beq
\psi_{\ell} \rightarrow \psi_{\ell} \, e^{i q_{\ell} \theta} \label{u1lutt}
\eeq
where $\theta$ generates the U(1) transformation, and $q_{\ell}$ is the charge of $\psi_{\ell}$.
Now the usual Noether argument can be used to generate an expression for the conserved charge density $\mathcal{Q}$.
This expression can depend upon specific details of the Lagrangian, and so we don't present a general form.
However, if the field $\psi_{\ell}$ is canonical, then its contribution to $\langle \mathcal{Q} \rangle$ is 
given by
\beq
\langle \mathcal{Q}_{\ell} \rangle = \pm \int \frac{d^d k}{(2 \pi)^{d}} \int_{-\infty}^{\infty} \frac{d \omega}{2\pi}  q_{\ell } G_{\ell} (k, i \omega) e^{i \omega 0^{+}} \label{ql}
\eeq
where $G_{\ell}$ is the 2-point Green's function of $\psi_{\ell}$, and leading sign refers to bosons/fermions. 
For simplicity, we will assume that suitable linear combinations of the fields can be chosen so that the Green's functions are diagonal. We emphasize that \eqnref{ql} applies only if $\psi_\ell$ is canonical, but the canonical nature is not required for the Luttinger theorem below in \eqnref{lutt}.

Now let us examine the dependence of $\langle \mathcal{Q} \rangle$ on an applied chemical potential $\mu$. The 
Noether argument implies that the Green's functions of all fields, whether canonical or not, depend upon $\mu$ only in the combination $G_{\ell} (k , \omega - q_{\ell} \mu)$; in other words, the chemical potential is merely a shift in the frequency.
Applying this shift to \eqnref{ql}, we might initially conclude that the frequency shift can be absorbed into a redefinition of the
dummy frequency variable which is being integrated over, and so the result is independent of $\mu$. However, this is not true
because the $\omega$ integration is only conditionally convergent at large $\omega$, and a finite result relies crucially on the
$e^{i \omega 0^{+}}$ convergence factor. The Luttinger theorem relies on an argument which extracts this conditionally convergent value, which turns out to be insensitive to many details of the Green's functions. The final result will yield the value of $\langle \mathcal{Q} \rangle$, whether or not the fields are canonical.

To proceed, it is useful isolate Feynman diagrams which are convergent at large $\omega$, and so do not require
the $e^{i \omega 0^{+}}$ convergence factor. In the evaluation of the free energy, such contributions are known
as the Luttinger-Ward functional \cite{LW,BK,chitra,pothoff} 
$Y_{LW}[ G_{\ell} (k, \omega)]$. This is a functional of all the fully renormalized
Green's functions, and is the sum of all closed-loop skeleton Feynman diagrams which are two-particle irreducible. 
The two-particle irreducibility ensures convergence at large $\omega$, and so the Luttinger-Ward functional obeys
\beq
Y_{LW}[ G_{\ell} (k, \omega - q_{\ell} \Omega)] = Y_{LW}[ G_{\ell} (k, \omega)] 
\eeq
for all $\Omega$. This identity is a key ingredient in the proof of the Luttinger result.
We now refer the reader to Ref.~\onlinecite{powell} (see Section V) and Ref.~\onlinecite{piers} (see Section IV) 
for further details, and proceed to the final result:
\beq
\langle \mathcal{Q} \rangle = \sum_{\ell \, \in \, {\rm fermions}} q_\ell V_\ell
\label{lutt}
\eeq
where $V_{\ell}$ is the momentum space volume enclosed by the Fermi surface of only the fermionic particles.
We emphasize that 
\begin{itemize} 
\item
$\mathcal{Q}$ measures the contribution to the total charge from both fermions and
bosons,\cite{powell} as determined by applying the Noether argument to the symmetry \eqnref{u1lutt}.
\item The sum on the right-hand-side of \eqnref{lutt} involves all fermions, whether canonical or not.\cite{piers}
\item Some of the Fermi surfaces could be of fermions which carry additional charges of fluctuating gauge fields, or are
coupled to other gapless scalars associated with a symmetry-breaking transition. 
In these cases, the singularities near the Fermi surface can differ from those in a Landau Fermi liquid. If the fermions have
gauge charges, then the fermion Green's functions and the singularities near the Fermi surface are gauge-dependent; however
the volume enclosed by the Fermi surface is gauge-independent. 
\item The Luttinger relation in Eq.~(\ref{lutt}) does not apply if the global U(1) symmetry is spontaneously broken, usually by
the condensation of a boson which carries charge $\mathcal{Q}$.
\end{itemize}

Let us now define the Fermi surface more precisely: it is the locus of points in momentum space where the inverse
Green's function has a zero at $\omega =0$. Assuming momentum space isotropy for simplicity, the Fermi momentum $k_{F}$
is defined by
\beq
G_{\ell}^{-1} (k = k_{F}, \omega = 0) = 0 \label{zero}
\eeq
as long as $\psi_{\ell}$ is a fermion. Unitarity conditions on the spectral representation do not allow bosons to satisfy \eqnref{zero} in general, because it would lead to instabilities of bosons over a range of momenta (see Ref.~\onlinecite{hong1} for a recent discussion of this); exceptions can arise at certain exotic critical points in spatially isotropic systems, which we will not consider here.
This is why there is no bosonic contribution to the right hand side of \eqnref{lutt}.
The proof of Luttinger's theorem also requires an additional mild condition on the fermion Green's functions:
\beq \lim_{\omega \rightarrow 0} \mbox{Im} \, G_{\ell}^{-1} (k \neq k_{F}, \omega) = 0, \label{imG} \eeq 
while $\mbox{Re} \, G_{\ell}^{-1} (k \neq k_{F}, 0) \neq 0$. 
In a Fermi liquid, the expression in \eqnref{imG} vanishes as $\omega^{2}$, but
this behavior is not needed for \eqnref{lutt}. The latter result applies for a much broader class of compressible ``non-Fermi liquid'' states, which usually have a slower approach to the $\omega \rightarrow 0$ limit.

As we will see below, many of the non-Fermi liquid states we find, and in particular those associated with couplings to deconfined gauge fields,
are ultimately unstable to paired superfluid states in which the global U(1) $\mathcal{Q}$ symmetry is broken. However, even in these cases
it is interesting to study the non-Fermi liquid ``normal'' state, because we can add additional perturbations that suppress the superfluidity. 
Moreover, the superfluidity may only appear at a low energy scale, and so there is a wide intermediate energy regime over
which the non-Fermi liquid physics applies.

The plan for the remainder of the paper is as follows. We will present a unified perspective on previously studied condensed matter models
in Sections~\ref{sec:doublon},~\ref{sec:bf}, and~\ref{sec:ffl}. Section~\ref{sec:doublon} will consider one of the simplest examples
of a compressible non-Fermi liquid phase, the doublon metal, which has 2 Fermi surfaces of fermions coupled to a U(1) gauge field with opposite charges. We study the phases of mixtures of bosons and fermions, without gauge fields, in Section~\ref{sec:bf}. Gauge fields are introduced to
Bose-Fermi mixtures in Section~\ref{sec:ffl}, and such models are connected to ``slave particle'' realizations of electronic Hubbard or Kondo models. This model has a fractionalized Fermi liquid (FL*) phase, which plays an important role in the connection to dual gravity models.
We turn to supersymmetric gauge theories, and the nature of the phase diagrams at non-zero chemical potential in 
Sections~\ref{sec:ABJM} and~\ref{sec:SYM} which describe the $d=2$ ABJM model and the $d=3$ $\mathcal{N}=4$ SYM models
respectively. We will find that the phases appearing in these models are closely connected to those discussed earlier in the condensed matter models. Finally, Section~\ref{sec:dis} presents a summary of our results.

{\em Note added:} Two complementary papers appeared just as the present paper was being submitted, addressing similar questions
and models from the dual gravity perspective. Ref.~\onlinecite{gauntlett} addressed the $d=2$ ABJM model of Section~\ref{sec:ABJM},
and Ref.~\onlinecite{aprile} addressed the $d=3$ SYM model of Section~\ref{sec:SYM}, both at non-zero chemical potentials.

\section{Doublon metal}
\label{sec:doublon}

Our simplest example of a non-Fermi liquid obeying the Luttinger theorem is the doublon metal.\cite{rkk,gs,ms} This is a model of 
a fluctuating doped antiferromagnet, with applications to the cuprate superconductors. The theory begins with an ordered
antiferromagnet and re-expresses the electrons in terms of the pocket Fermi surfaces created by the antiferromagnetic order.
The antiferromagnetic moment is then allowed spacetime fluctuations in orientation, and these lead to an emergent U(1) 
gauge field which is coupled to the electron pockets. The simplest form of such a theory has a pair of fermions,
$f_\pm$; these are sometimes called `doublons' because they represent doubly-occupied sites in a derivation from 
a lattice Hubbard model (the doublons were denoted $g_{\pm}$ in the earlier work\cite{rkk,gs,ms}). These doublons are
coupled with opposite charges to the U(1) gauge field $(A_\tau, {\bf A})$, as described by the 
Lagrangian\cite{gs,ms,wenholon,schulz,leeholon,shankar}
\begin{eqnarray}
\mathcal{L}_d &=& f_+^\dagger \left[ (\partial_\tau - i A_\tau )  - \frac{( {\bm \nabla} - i {\bf A}  )^2}{2m_f}  - \mu  \right] f_+
\nonumber \\ &+& f_-^\dagger \left[ (\partial_\tau + i A_\tau )  - \frac{( {\bm \nabla} + i {\bf A}  )^2}{2m_f} - \mu \right] f_- . \label{lg}
\end{eqnarray}
We have not written out a possible bare Maxwell term for the gauge field, because it is irrelevant at low energies compared to
the contributions of the fermion polarization.

\begin{figure}[h]
\begin{center}
\includegraphics[width=1in]{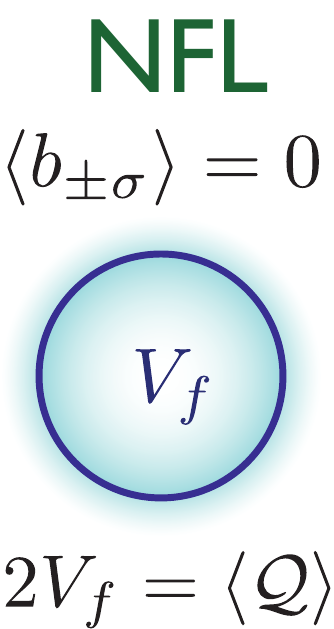}
\end{center}
\caption{The non-Fermi liquid (NFL) doublon metal phase. The blue blurry shading of the Fermi surface indicates the coupling of the 
$f_\pm$ fermions to a fluctuating gapless gauge field, so that the fermion Green's function has the singular behavior of Eq.~(\ref{nfl}) near
the Fermi surface (in $d=2$). Despite the non-Fermi liquid character of the fermion excitations, the value of $k_F$, and so the 
location of the Fermi surface, is sharply defined. The global SU(2) spin is carried by the bosons $b_{\pm \sigma}$ which are gapped
in this phase.}
\label{pd_nfl}
\end{figure}

This theory has a global U(1) charge 
\beq
\mathcal{Q} = f_+^\dagger f_+ + f_-^\dagger f_-, 
\eeq
while the gauge field
couples to the orthogonal charge $f_+^\dagger f_+ - f_-^\dagger f_-$. By the Luttinger theorem discussed in 
\secref{sec:lutt}, a compressible phase with the global U(1) unbroken must have Fermi surfaces of the $f_\pm$ fermions.
Because of the interchange symmetry between the fermions, the two Fermi wavevectors must be equal, and so the
Fermi volume $V_f$ of each Fermi surface obeys 
\beq
2 V_f = \left \langle \mathcal{Q} \right \rangle.
\eeq
We have sketched a pictorial representation of this non-Fermi liquid (NFL) phase in Fig.~\ref{pd_nfl}.

There is a long history of studies on the influence of the gauge field fluctuations on such Fermi surfaces. While the longitudinal gauge
field fluctuations are screened, the transverse fluctuations lead to singular non-Fermi liquid renormalizations of the fermions
near the Fermi surface. Such fluctuations are frequently controlled via a $1/N$ expansion, where each fermion is endowed with an additional flavor index which can take $N$ values. Recent work \cite{leen,metnem,mross} has shown 
that the naive $1/N$ expansion breaks
down in $d=2$ because of Fermi surface singularities that appear in higher loop graphs. The $d=2$ case is therefore strongly-coupled, and the ultimate fate of the theory has not been fully resolved: these are difficult questions we will not address here.

Despite the strong-coupling nature of the problem, the recent studies do point to a natural scaling structure for the fermion Green's function in the vicinity of the Fermi surface. We focus on the $d=2$ case, and in vicinity of any point, say ${\bf k}_0 = (k_F, 0)$
on the Fermi surface. Then, we measure the fermion momentum, ${\bf k}$, using deviations from this point
\beq {\bf q} = {\bf k} - {\bf k}_0. \eeq
The singularity in the fermion Green's function scales as a function of the distance to the nearest point on the Fermi surface, which is 
\beq
q \equiv |{\bf k}| - k_F \approx q_x + \frac{q_y^2}{2 k_F}; \eeq
note that we have to scale $q_x \sim q_y^2$ as we approach the Fermi surface. The vicinity of the Fermi surface in the 
doublon metal is described by \cite{metnem}
\beq
G^{-1} ({\bf k}, \omega) = q^{1-\eta} \Phi (\omega/q^{z/2}) \label{nfl}
\eeq
where $\eta$ and $z$ are anomalous exponents and $\Phi$ is a scaling function 
which can be computed at low orders in the $1/N$ expansion. The structure of $\Phi$ is such that the relations in
\eqnref{zero} and \eqnref{imG} are obeyed, and so the Luttinger theorem does apply in the doublon metal.

Note that the Green's function in \eqref{nfl} is gauge-dependent, and computations are normally made in the 
Coulomb gauge ${\bm \nabla} \cdot {\bf A} = 0$. However, the Fermi surface can also leave its fingerprints in correlations
of gauge-invariant observables. A prominent example is the two-point correlator of the density $\mathcal{Q}$, which would
have spatial oscillations at the wavevector $2 k_F$, which are analogs of the Friedel oscillations of Fermi liquids.

An important feature of the doublon metal is that it has an instability towards superconductivity via the appearance
of a condensate of the Cooper pair $f_+ f_-$: this is a consequence of the attractive interaction between the $f_+$
and $f_-$ fermions mediated by the gauge field. The Cooper pair is gauge neutral, and so the gauge symmetry remains
unbroken in the state with $\langle f_+ f_- \rangle \neq 0$. However, such a condensate does 
break the global U(1) symmetry associated with $\mathcal{Q}$. So the conditions on the Luttinger theorem are not obeyed,
and there is no constraint on the Fermi surface volume in the superconducting state. Indeed, in the present model, the Fermi surfaces are immediately
gapped by any non-zero condensate.

The existence of the doublon metal therefore requires that the pairing scale be suppressed to a very low energy so that
there is a significant intermediate energy scale for non-Fermi liquid physics. Clearly, a large bare repulsive interaction 
between the fermions can help establish such a regime. Determining the precise conditions and width of a possible doublon metal regime involve strong-coupling questions which will be addressed in a forthcoming paper. 
Such a pairing instability is an ``affliction'' common to many of the other non-Fermi liquid compressible
phases we will consider in the present paper.

For completeness, we also note the structure of the spin excitations of the doublon metal; the reader can skip the remainder of this
section without loss of continuity. Here ``spin'' refers to a global SU(2) symmetry of 
lattice electronic models like the Hubbard model, and is analogous to global ``flavor'' symmetries in relativistic field theories.
The spinful excitations are bosons $b_{\pm \sigma}$, which carry the charge, $\pm 1$, of the U(1) gauge field $(A_\tau, {\bf A})$,
along with the global spin quantum number $\sigma = \uparrow, \downarrow$. However, these bosons do not carry the global U(1)
charge $\mathcal{Q}$. The bosons have an energy gap, and their low energy excitations are described by the relativistic CP$^1$
model \cite{rkk}; however we write it here in a non-relativistic notation, to highlight the connections to models to be considered later in this paper:
\bea
\mathcal{L}_\sigma &=&  b_{+\sigma}^\dagger \left[ (\partial_\tau - i A_\tau )  - \frac{( {\bm \nabla} - i {\bf A}  )^2}{2m_b} + \epsilon_1  \right] b_{+\sigma}
\nonumber \\ &+& b_{-\sigma}^\dagger \left[ (\partial_\tau + i A_\tau )  - \frac{( {\bm \nabla} + i {\bf A}  )^2}{2m_b} + \epsilon_1 \right] 
b_{-\sigma} + \epsilon_2 \left( \varepsilon_{\sigma\sigma'} b_{+\sigma} b_{- \sigma'} + \hc \right),
\label{lsigma}
\eea
where $\varepsilon_{\sigma\sigma'}$ is the unit antisymmetric tensor.
The bosonic spin excitations have an energy gap $\sqrt{\epsilon_1^2 - \epsilon_2^2}$, as is easily seen by diagonalizing the quadratic
form of $\mathcal{L}_\sigma$.

The combined theory $\mathcal{L}_d + \mathcal{L}_\sigma$ has a number of possible phases, distinct from the 
doublon metal \cite{rkk,naturephys,su2,yang,fflsc}. Condensation of $b_{\pm \sigma}$ breaks the U(1) gauge symmetry and
leads to antiferromagnetic order. More interesting for our purposes here are phases associated with the formation of gauge-neutral
composites of $b_{\pm \sigma}$ and $f_{\pm}$, yielding `electron'-like fermions $c_\sigma$ which can have their own Fermi 
surfaces.
Rather than discussing such phases here, we will examine analogous phases in a simpler model in the following section,
and also find several similar phases in the phase diagrams of the models of subsequent sections.

\section{Boson-fermion mixture}
\label{sec:bf}

Now consider a quantum liquid which is a mixture of fermions, $f_\sigma$, 
and bosons $b$. As in Section~\ref{sec:doublon}, $\sigma$ is a global SU(2) spin `flavor' index, and it does
not play a significant role in this section. We could drop the $\sigma$ index below, but we retain
it because of its importance in physical analogies to be discussed later.

A common physical example is a mixture of $^3$He and $^4$He. Other examples have been studied
recently in ultracold trapped atom systems, such as $^6$Li and $^7$Li.
With weak interactions between the bosons and fermions, each proceed relatively independently. 
The bosons condense to form a superfluid (SF), while the fermions from a Fermi liquid (FL), with the volume enclosed by
the Fermi surface equal to the fermion density.

Now we turn up the interaction strength between the $f_\sigma$ and 
$b$, so that in free space a single $f_\sigma$ and $b$ can bind to
form a fermionic molecule. We are interested here in the consequences of this 2-body physics for the 
many body problem. A focus on the Fermi surfaces, and the Luttinger theorem, allows us to make sharp
distinctions between phases in the case of a dense gas.\cite{powell}

Let us write a simple Lagrangian which can describe this physics:
\bea
\mathcal{L}_{bf} &=& f_\sigma^\dagger \left[ \partial_\tau   - \frac{{\bm \nabla}^2}{2 m_f} - \mu  \right] f_\sigma + 
b^\dagger \left[ \partial_\tau   - \frac{{\bm \nabla}^2}{2 m_b} - \mu_b  \right] b \nn
&~& \quad \quad + \frac{u}{2} \left(b^{\dagger} b \right)^{2}- g f_\sigma^\dagger b^\dagger b f_\sigma \label{Lbf}
\eea
This theory clearly has 2 global U(1) symmetries, with conserved charges
\bea {\rm U(1)}&:& \quad \mathcal{Q} = f_\sigma^\dagger f_\sigma \nn 
\mbox{U$_b$(1)}&:& \quad\mathcal{Q}_b = b^\dagger b, \label{Qfb}
\eea
and chemical potentials $\mu$ and $\mu_b$ coupling to these charges. There is a repulsive interaction $u>0$ between the bosons
necessary to stabilize the theory, and an attractive interaction $g$ between the bosons and fermions.
Now it is useful to introduce a fermionic `molecular' field $c_\sigma$ by  a Hubbard-Stratonovich decoupling \cite{piers} 
of the two-body interaction:
\bea
\mathcal{L}_{bf} &=& f_\sigma^\dagger \left[ \partial_\tau   - \frac{{\bm \nabla}^2}{2 m_f} - \mu  \right] f_\sigma + 
b^\dagger \left[ \partial_\tau   - \frac{{\bm \nabla}^2}{2 m_b} - \mu_b  \right] b \nn
&~& \quad \quad + \frac{u}{2} \left(b^{\dagger} b \right)^{2} + \frac{1}{g} c_\sigma^\dagger c_\sigma - c_\sigma^\dagger  b f - f^\dagger b^\dagger c_\sigma . \label{Lbfc}
\eea
Note that the field $c_\sigma$ transforms under both U(1) symmetries in \eqnref{Qfb}, which are now associated respectively with
\bea
{\rm U(1)}&:& \quad f_\sigma \rightarrow f_\sigma \, e^{i \theta}, \quad c_\sigma \rightarrow c_\sigma \, e^{i \theta} \nn
\mbox{U$_b$(1)}&:& \quad b \rightarrow b \, e^{i \theta_b}, \quad c_\sigma \rightarrow c_\sigma \, e^{i \theta_b}. \label{UU}
\eea
However the field $c$ is not canonical, and application of the Noether argument to \eqnref{Lbfc} shows that the expressions
for the charge $\mathcal{Q}$ and $\mathcal{Q}_b$ in \eqnref{Qfb} remain {\em unchanged\/}.

\begin{figure}
\begin{center}
\includegraphics[width=6in]{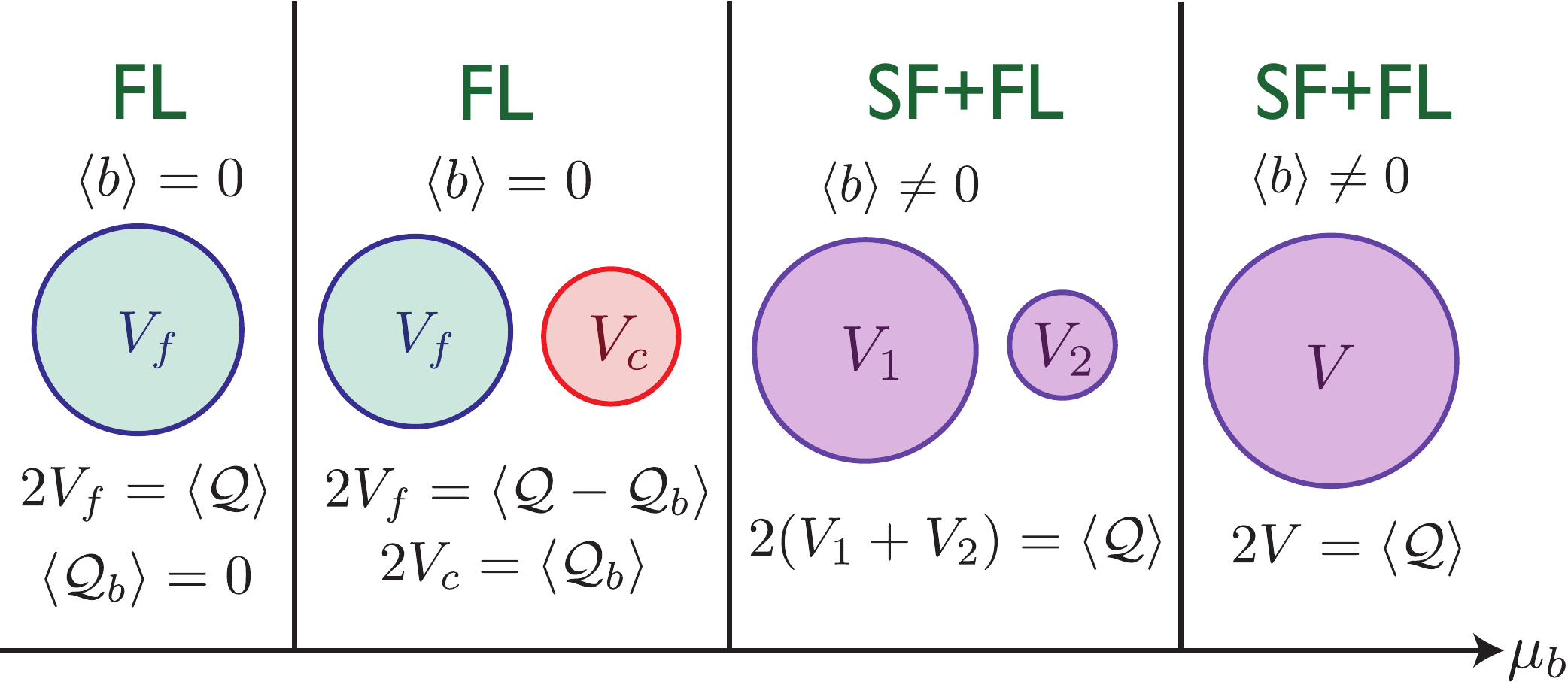}
\end{center}
\caption{Schematic phase diagram of the theory $\mathcal{L}_{bf}$ in Eq.~(\ref{Lbf}) for a strong interaction $g$; 
with weak interactions, the two intermediate phases are not present---see Ref.~\onlinecite{powell}
for more details.
The Fermi liquid (FL) phases have no Bose condensate, and the two global U(1) symmetries constrain the two
Fermi surfaces of the $f_\sigma$ and $c_\sigma$ fermions via the Luttinger relation in Eq.~(\ref{Vcf}). 
Unlike the model of Section~\ref{sec:doublon}, the Fermi surface excitations are not coupled to a fluctuating
gauge field, and Fermi liquid-like quasiparticles survive near the Fermi surface; this is indicated by the uniform
shading within the Fermi surface. The case with only a $c_\sigma$ Fermi surface is allowed only for $\left\langle Q \right\rangle = \left\langle Q_b \right\rangle$.
The SF+FL phases have both a Bose condensate and Fermi surfaces; the non-zero $\langle b \rangle$
hybridizes the $f_\sigma$ and $c_\sigma$ fermions, the Fermi surface quasiparticles are therefore linear
combinations of $f_\sigma$ and $c_\sigma$. There can be one or two such Fermi surfaces as shown above,
depending upon parameters. There is only one Luttinger constraint on the volumes of the Fermi surfaces in 
the SF+FL phases. Here and in the following figures, we follow the convention of shading $f_\sigma$ Fermi surfaces blue, $c_\sigma$ Fermi surfaces red,
and Fermi surfaces of hybridized fermions purple. 
 }
\label{pd_bf}
\end{figure}

This theory can now have distinct phases, depending upon whether U$_b$(1) is broken or not, as is shown 
in Fig.~\ref{pd_bf}. 
In all phases, Fermi surfaces
of both the $f_\sigma$ and $c_\sigma$ fermions can be present: let these enclose volumes $V_f$ and $V_c$ respectively. 
By \eqnref{lutt} and \eqnref{UU}, the volume $V_c$ will be included in the Luttinger count, even though $c$ does not appear
in the expressions for the conserved densities in \eqnref{Qfb}. 

The Fermi liquid (FL) phases have no $b$ condensate, and both global symmetries are preserved. 
Then by the Luttinger theorem, there have to be 2 restrictions on the Fermi volumes. From \eqnref{lutt} and \eqnref{UU} these are easily seen to be \cite{powell}
\bea
\langle \mathcal{Q} \rangle &=& \langle f_\sigma^\dagger f_\sigma \rangle = 2 (V_f + V_c) \nn 
\langle \mathcal{Q}_b \rangle &=& \langle b^\dagger b \rangle =  2V_c , \label{Vcf}
\eea
where the factors of 2 arise from the sum over $\sigma$.
Thus both $V_f$ and $V_c$ are fixed by the densities of the underlying bosons and fermions. Remarkably, the volume $V_c$ is constrained by the number of bosons: intuitively, this means that all the bosons have to bind with a $f_\sigma$ fermion to form 
a fermionic molecule $c_\sigma$ to avoid Bose condensation. The case $\left\langle \mathcal{Q}_b \right\rangle = 0$ has $V_c=0$
and hence only a $f_\sigma$ Fermi surface.
Similarly,
the case with $\left\langle Q \right\rangle = \left\langle Q_b \right\rangle$
has $V_f=0$ and only a $c_\sigma$ Fermi surface. It is not possible to have $\left\langle Q \right\rangle < \left\langle Q_b \right\rangle$
in a FL phase.

The other phases are where $b$ condenses and U$_b$(1) is broken; such phases include the region where
$\left\langle Q \right\rangle < \left\langle Q_b \right\rangle$.
 Now only the first of the Luttinger constraints in
\eqnref{Vcf} applies. This is clearly the same as the superfluid (SF) state discussed at the beginning of this section.
The superfluid order co-exists with Fermi surfaces of the fermions, and depending upon the magnitude of 
$\langle b \rangle$ and other parameters, there can be one or two Fermi surfaces as shown in Fig.~\ref{pd_bf}.
A SF only phase is only possible when $\langle \mathcal{Q} \rangle = 0$, {\em i.e.\/} there are no fermions.

\section{Fractionalized Fermi liquid}
\label{sec:ffl}

The fractionalized Fermi liquid (FL*) was introduced in Refs.~\onlinecite{ffl1,ffl2} as a compressible non-Fermi liquid
phase of Kondo  and Hubbard lattice models of strongly interacting electrons. Here, we will introduce the FL* in the context
of continuum field theories of fermions and bosons under consideration. See Ref.~\onlinecite{tasi} for a review of the connection to these condensed matter lattice models. 

Here, we begin with the model of \secref{sec:bf} and gauge the U(1) charge $\mathcal{Q} - \mathcal{Q}_b$. 
Thus we have a dynamic U(1) gauge field $(A_\tau, {\bf A})$ (as in \secref{sec:doublon}), and the Lagrangian in
\eqnref{Lbf} is modified to 
\bea
\mathcal{L}_{*} &=& f_\sigma^\dagger \left[ (\partial_\tau - i A_\tau)   - \frac{({\bm \nabla}- i {\bf A})^2}{2 m_f} - \mu  \right] f_\sigma 
\nn &+& 
b^\dagger \left[ (\partial_\tau + i A_\tau)   - \frac{({\bm \nabla} + i A)^2}{2 m_b} - \mu_b  \right] b \nn
 &~& \quad \quad + \frac{u}{2} \left(b^{\dagger} b \right)^{2} - g f_\sigma^\dagger b^\dagger b f_\sigma + i A_\tau \rho . \label{Ls}
\eea
The last term is a background charge density $\rho$: this is needed here because stability requires that a U(1) gauge field
only interact with matter which has net zero U(1) charge density. So we must have
\beq
\rho = \left\langle \mathcal{Q} \right\rangle - \left\langle \mathcal{Q}_b \right\rangle, \label{defrho}
\eeq
where the definitions of the charges are just as in \eqnref{Qfb}.
We can introduce the composite field $c_\sigma$ just as in \eqnref{Lbfc}: this field is gauge-invariant.

\begin{figure}
\begin{center}
\includegraphics[width=6in]{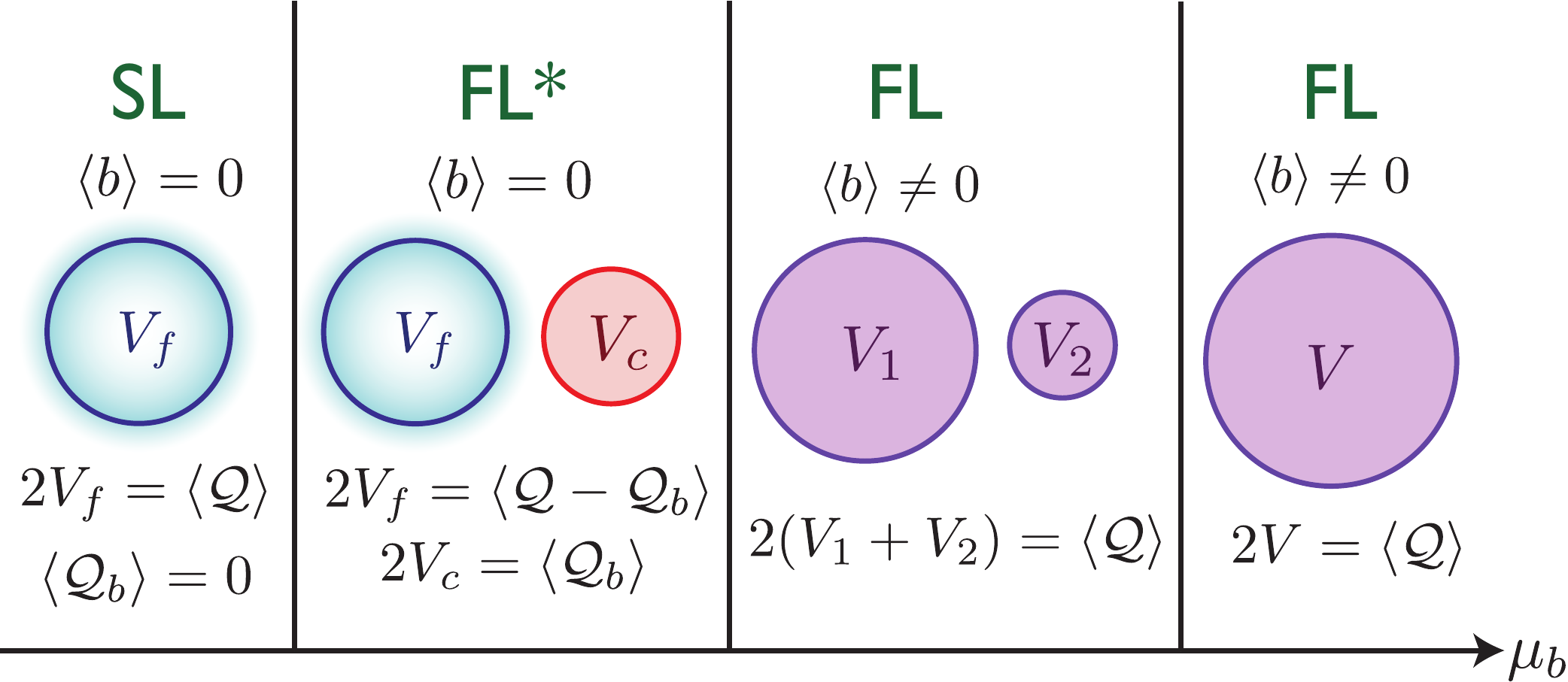}
\end{center}
\caption{Schematic phase diagram of the theory $\mathcal{L}_{*}$ in Eq.~(\ref{Ls}).
This is similar to the phase diagram in Fig.~\ref{pd_bf}, but some of the Fermi surface excitations are
now coupled to a fluctuating gapless gauge field: such Fermi surfaces are indicated by the blurry shading, as in Fig.~\ref{pd_nfl}.
The colors of the Fermi surfaces are chosen as in Fig.~\ref{pd_bf}.
The fractionalized Fermi liquid (FL*) phase has Fermi surfaces of both gauge-neutral and gauge-charged fermions. The spin liquid (SL) 
phase has only a gauge-charged Fermi surface and
is incompressible, and is the only incompressible phase in our phase diagrams.
Unlike Fig.~\ref{pd_bf}, the phases with a $b$ condensates are not superfluids because there is no gauge-invariant condensate which
violates a global U(1) conservation.
}
\label{pd_ffl}
\end{figure}

The phases of $\mathcal{L}_{*}$ closely parallel those of $\mathcal{L}_{bf}$: the main difference is that the gauge fluctuations
can modify the nature of the singularities near the Fermi surfaces. A schematic phase diagram appears in Fig.~\ref{pd_ffl}.

The FL* phase is obtained when $b$ is uncondensed, and the U(1) gauge theory is in a deconfined phase. There are both 
$f_\sigma$ and $c_\sigma$ Fermi surfaces, and their volumes continue to obey both constraints in \eqnref{Vcf}. 
Now the gauge fluctuations will lead to singularities on the $f_\sigma$ Fermi surfaces described by \eqnref{nfl}. The $c_\sigma$ Fermi surface 
involves gauge-invariant fermions, and so has weaker singularities; however they will not be Fermi liquid-like because
the $c_\sigma$ fermions do couple to the $f_\sigma$ sector, albeit only through gauge-invariant operators.

An important point is that only the $c_\sigma$ Fermi surface is observable as a sharp resonance in the
spectral resonance of fermions which carry charge $\mathcal{Q}$; such resonances are detected in photoemission
experiments in the context of the condensed matter models. This is because these probes only detect gauge-invariant operators. 
Thus such probes will see
a deficit in the Luttinger count, as the observed Fermi volume will not equal the total fermion density $\langle \mathcal{Q} \rangle$.
In reality the full Luttinger count in \eqnref{Vcf} is obeyed, and is made up by ``hidden'' Fermi surfaces of the gauge-dependent 
fermions $f_\sigma$; these hidden Fermi surfaces only appear as weaker singularities in gauge-invariant observables, as we discussed in
Section~\ref{sec:doublon}.
This
deficit in the observed Fermi volume is a key characteristic of the FL* phase.

A special case of the FL* phase is the spin liquid 
SL phase (see Fig.~\ref{pd_ffl}) which has $\left\langle \mathcal{Q}_b \right\rangle = 0$,
and so there is no gauge-neutral Fermi surface.
Because of Eq.~(\ref{defrho}), the SL phase also has fixed $\langle \mathcal{Q} \rangle$ and so is incompressible: it is the only incompressible
phase we consider. In the application to lattice Kondo  or Hubbard  models, the SL phase is an insulator.

Finally, the remaining phases of $\mathcal{L}_*$ are the Higgs phases where $b$ condenses and gaps out the gauge fluctuations.
We can also view these as confining phases of the U(1) gauge theory, because of the continuity of confinement with Higgs phases
of fundamental scalars \cite{peskin}.
These are ordinary Fermi liquids (FL), and only the first constraint in \eqnref{Vcf} applies to Fermi volumes. 
As in Section~\ref{sec:bf}, the $f_\sigma$ and $c_\sigma$ fermions hybridize via the $b$ condensate, and lead to 
Fermi surfaces with ordinary Fermi liquid-like singularities. Depending upon parameters, there can be one or two such Fermi surfaces,
as indicated in Fig.~\ref{pd_ffl}. Note that, unlike Fig.~\ref{pd_bf}, there is no SF order in the FL phases. This is because
now the $b$ condensate carries a gauge charge, and there is no gauge-invariant condensate which breaks a global U(1) symmetry.

The reader is referred to Ref.~\onlinecite{tasi} for a review of the charge transport properties of these phases.

\section{Theory similar to the ABJM model in $d=2$}
\label{sec:ABJM}

This section will extend our study of compressible quantum matter to the canonical model
of AdS/CFT duality in $d=2$ spatial dimensions: the ABJM theory.\cite{abjm}
This gauge theory has $\mathcal{N}=6$ supersymmetry along with a global
SU(4) symmetry.

Here we will move away from the superconformal fixed point by adding a chemical potential
which couples to one of the generators of SU(4). This induces unstable directions in the potential
of the scalar fields in the theory, and so it seems that such a deformation may not be well defined.
However, it should be noted that the chemical potential also greatly reduces the symmetry of 
theory: it breaks supersymmetry and also reduces the SU(4) global symmetry. Thus, 
we should allow additional terms in the effective action, consistent with the reduced symmetry.
It seems plausible that these additional terms can be chosen to render the theory stable.
This section will present a simple toy model which can capture the possible compressible phases
of such a stable theory.

Benna {\em et al.}\cite{benna} have given an explicit formulation of the ABJM theory
which is suitable for our purposes: see their Section 4. The theory has two-component Dirac
fermions and complex scalars, both of which are bi-fundamentals of a U($N$)$\times$U($N$) gauge
group, and fundamentals of the global SU(4) `flavor' symmetry. Using the notation of Eqs. (4.21) and (4.23) 
of Benna {\em et al.}\cite{benna}, we choose the SU(4) generator $\mbox{diag}(1,1,-1,-1)$
as our global U(1) charge $\mathcal{Q}$. 
We perform a particle-hole transformation on 
particles on the bottom 2 components, and so then all particles carry a unit U(1) global charge.
In the presence of such a chemical potential, there is a residual SU(2)$\times$SU(2) flavor symmetry;
we will drop this flavor symmetry for simplicity.
Also, we will work with non-relativistic particles which only carry charges favored by the chemical potential.
Finally, we will reduce the U($N$)$\times$U($N$) gauge group to the simplest possible U(1) gauge group.

We note that most of the simplifications above are not essential. We just wish to work in the simplest
possible model, and our analysis below can be easily extended to include the features we have deemed inessential so far.
In particular, other choices for the generator of the global U(1) charge $\mathcal{Q}$ lead to similar results.

By this reasoning, we end up with 2 species of non-relativistic fermions $f_{+}$ and $f_{-}$, which
carry opposite charges under the U(1) gauge group; the negative  gauge-charged particles were obtained when
we performed the particle-hole transformation to obtain positive global U(1) charges above.
We also have a U(1) gauge field $(A_{\tau}, {\bf A})$.
Remarkably, so far the particle and gauge-field content, and global and gauge symmetries 
are {\em identical\/} to the theory $\mathcal{L}_d$ in Eq.~(\ref{lg}) of the doublon metal in \secref{sec:doublon}.
In addition, the present model also has bosons, $b_{+}$ and $b_{-}$, which carry the same
gauge and global charges as the fermions: the model for these bosons differs from $\mathcal{L}_\sigma$ in 
Eq.~(\ref{lsigma}), because the bosons of Section~\ref{sec:doublon} do not carry the global U(1) charge $\mathcal{Q}$.
Instead, the boson sector is similar to that of complementary theories of doped antiferromagnets \cite{wenlee,balents1,balents2} 
with a different pattern of electron fractionalization.

We can now write down a Lagrangian guided by the structure of the ABJM model\cite{benna}, or equivalently.
using the strategies of Sections~\ref{sec:doublon} and~\ref{sec:ffl}. The ABJM model has a large number of
quartic couplings between the fermions and bosons, but first we only include those which convert a pair of bosons into
a pair of fermions: these terms will be important for the structure of our mean-field theory. 
Thus our Lagrangian is, so far
\bea
\mathcal{L}_0 &=& f_+^\dagger \left[ (\partial_\tau - i A_\tau )  - \frac{( {\bm \nabla} - i {\bf A}  )^2}{2m_f}  - \mu  \right] f_+
\nonumber \\ &+& f_-^\dagger \left[ (\partial_\tau + i A_\tau )  - \frac{( {\bm \nabla} + i {\bf A}  )^2}{2m_f} - \mu \right] f_- \nn
&+& b_+^\dagger \left[ (\partial_\tau - i A_\tau )  - \frac{( {\bm \nabla} - i {\bf A}  )^2}{2m_b} + \epsilon_{1} - \mu  \right] b_+
\nonumber \\ &+& b_-^\dagger \left[ (\partial_\tau + i A_\tau )  - \frac{( {\bm \nabla} + i {\bf A}  )^2}{2m_b} + \epsilon_{1} - \mu \right] b_- \nn
& +&  \frac{u}{2} \left(b_{+}^{\dagger} b_{+} +  b_{-}^{\dagger} b_{-}\right)^{2} + v \, b_+^\dagger b_-^\dagger b_- b_+ - g_{1} \, \left( b_{+}^{\dagger} b_{-}^{\dagger} f_{-} f_{+}  + \hc \right).
\label{l0a}
\eea
Here $\epsilon_{1}$ is a parameter which can be tuned to modify the relative densities of fermions and bosons, and will help
access different phases of our phase diagram.

We can add additional quartic interactions between the fermions and the bosons, but we will decouple them by
a Hubbard-Stratonovich transformation using a gauge-invariant fermion $c$, as in Sections~\ref{sec:bf} and~\ref{sec:ffl}.
Unlike these earlier sections however, here we
will make fermion $c$ canonical. This amounts to choosing slightly different short-distance physics, but helps access more phases already
in mean-field theory. We ultimately expect a similar phase diagram if $c$ was chosen non-canonical as in \secref{sec:ffl},
but after including loop corrections to the $c$ fermion self energy.
An analogy with the Fermi gas at unitarity helps clarify this point: this theory contains a composite boson which 
can be either canonical or non-canonical, and the choice only distinguishes different perturbative expansions of the same physics.\cite{nishidason,nikolic} Also, with attractive gauge forces between the $b\pm$ bosons and $f_\pm$ fermions, we can expect that they 
form multiple bound states, and each of these lead to separate Fermi surfaces: in particular, this expected from the multiple Fermi surfaces 
seen in a recent holographic analysis \cite{sean2,sean3}. For simplicity, we will only consider a single such bound state, and a single $c$ fermion here,
but it is not difficult to extend our analysis to multiple $c$ fermions.

Including the gauge-neutral canonical fermion $c$, our final form for the theory analogous to the ABJM model is 
\bea
\mathcal{L}_{1} &=& \mathcal{L}_{0} + \mathcal{L}_{c} \nn
\mathcal{L}_c &=& c^\dagger \left[ \partial_\tau  - \frac{{\bm \nabla}^2}{2m_c}  + \epsilon_{2} -2 \mu  \right] c
- g_{2} \left[ c^{\dagger} \left(f_{+}b_{-} + f_{-}b_{+} \right) + \hc \right]. \label{l1}
\eea
Here $\epsilon_{2}$ is another tuning parameter for the phase diagram. The fermion-boson coupling above
respects the discrete $Z_2$ symmetry $f_{\pm} \rightarrow i f_{\mp}$, $b_{\pm} \rightarrow b_{\mp}$, $c \rightarrow i c$, 
and $(A_{\tau}, {\bf A}) \rightarrow - (A_{\tau}, {\bf A})$ of the theory.
So we now have presented our complete theory $\mathcal{L}_{1}$ with a U(1) gauge invariance and a U(1) global symmetry; the latter has 
conserved charge
\beq
\mathcal{Q} = f^{\dagger}_{+} f_{+} + f^{\dagger}_{-} f_{-} + b^{\dagger}_{+} b_{+} + b^{\dagger}_{-} b_{-} + 2 c^{\dagger}c.
\label{Qabjm}
\eeq

As we have already noted, the theory $\mathcal{L}_0$ is remarkably similar to the theory $\mathcal{L}_d + \mathcal{L}_\sigma$
of the doublon metal in Section~\ref{sec:doublon}: the main difference is that bosons of the doublon metal do not carry
the global U(1) charge $\mathcal{Q}$, but have an additional global SU(2) spin (flavor) quantum number.
Instead, the boson and gauge sector of $\mathcal{L}_0$ is closely related to theories of doped antiferromagnets \cite{wenlee,balents1,balents2} in which the 
global charge is carried by the bosons, and the spin is carried by the fermions. Thus the theory $\mathcal{L}_0$ fragments into pieces equivalent to 
different models of doped quantum antiferroments, but there is no such precise correspondence for all of $\mathcal{L}_0$.

\subsection{Phase diagram}
\label{sec:pdabjm}

In the simplest mean-field theory of $\mathcal{L}_{1}$, we treat $b_{\pm}$ as $c$-numbers and ignore the gauge field. 
This will allow us
to determine qualitative aspects of the phase diagram, and we will subsequently discuss the full structure of the various
phases.
\begin{figure}[h]
\begin{center}
\includegraphics[width=4in]{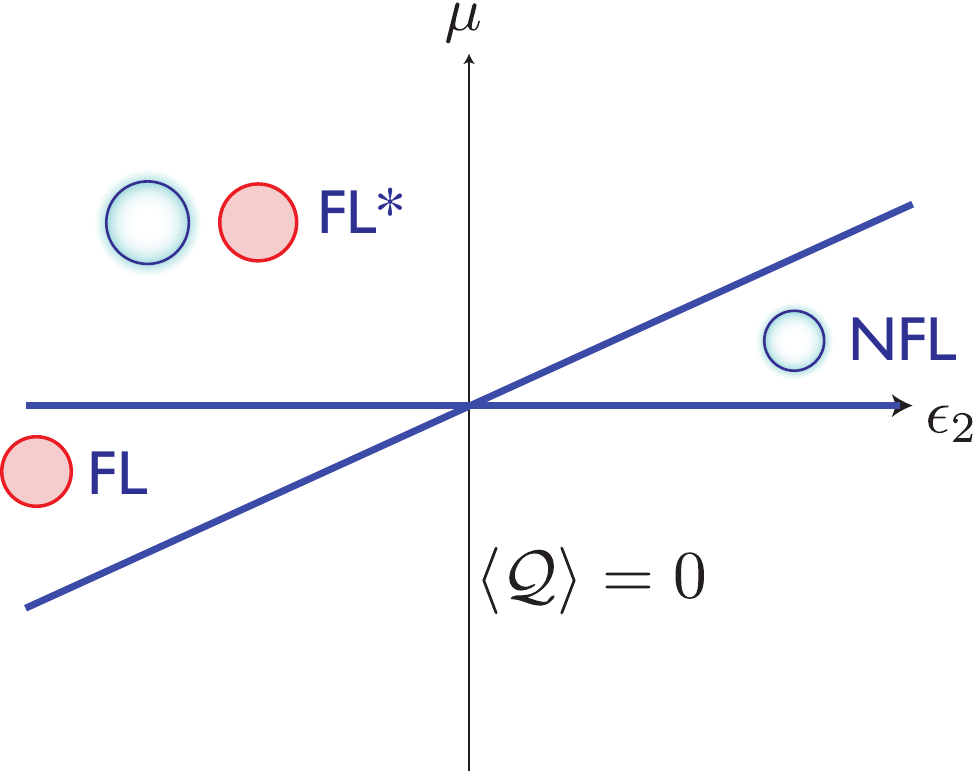}
\end{center}
\caption{Mean field phase diagram of the theory $\mathcal{L}_1$ in Eq.~(\ref{l1}) in the limit of very large $\epsilon_1$,
when we have $\langle b_\pm \rangle =0$, and there is no possibility of SF order. All phases are compressible, the global
U(1) symmetry is preserved, and the phases are distinguished by the configurations of the Fermi surfaces. The phase boundaries in this limit are at $\epsilon_2=2\mu$ and $\mu=0$.
The Fermi surfaces are colored as in Fig.~\ref{pd_ffl}. Fermi surfaces whose volumes are degenerate by symmetry are shown by
a single circle, while inequivalent Fermi surfaces are shown separately.  }
\label{pd_abjm0}
\end{figure}
\begin{figure}[h]
\begin{center}
\includegraphics[width=5in]{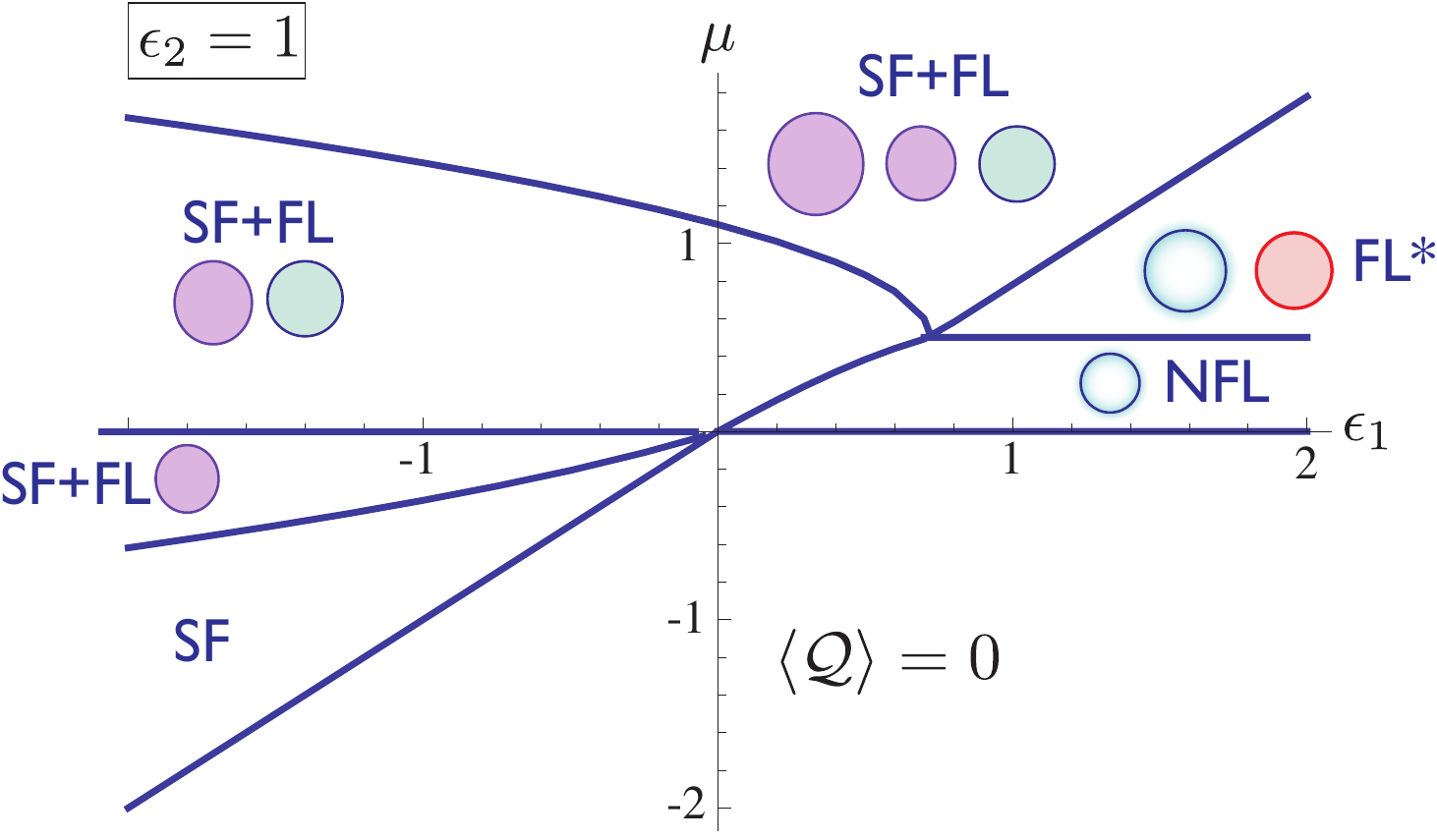}
\end{center}
\caption{The phase diagram of $\mathcal{L}_1$ for $g_1=0$ and $g_2 = 1$. 
The other parameters are shown, or described in the text. 
All the phases labeled SF have $\langle b_\pm \rangle \neq 0$, while the remainder have $\langle b_\pm \rangle =0$.
The Fermi surfaces are colored as in Fig.~\ref{pd_ffl} and~\ref{pd_abjm0}. For $g_1 =0$, all but the $v$ term in the energy depend only upon $|b_+|^2 + |b_-|^2$; we have assumed a small $v < 0$, so that degeneracy of the condensate is lifted, and we
have $\langle b_+ \rangle = \langle b_- \rangle \neq 0$ in all the phases with a SF label. }
\label{pd_abjm1}
\end{figure}
\begin{figure}[h]
\begin{center}
\includegraphics[width=5in]{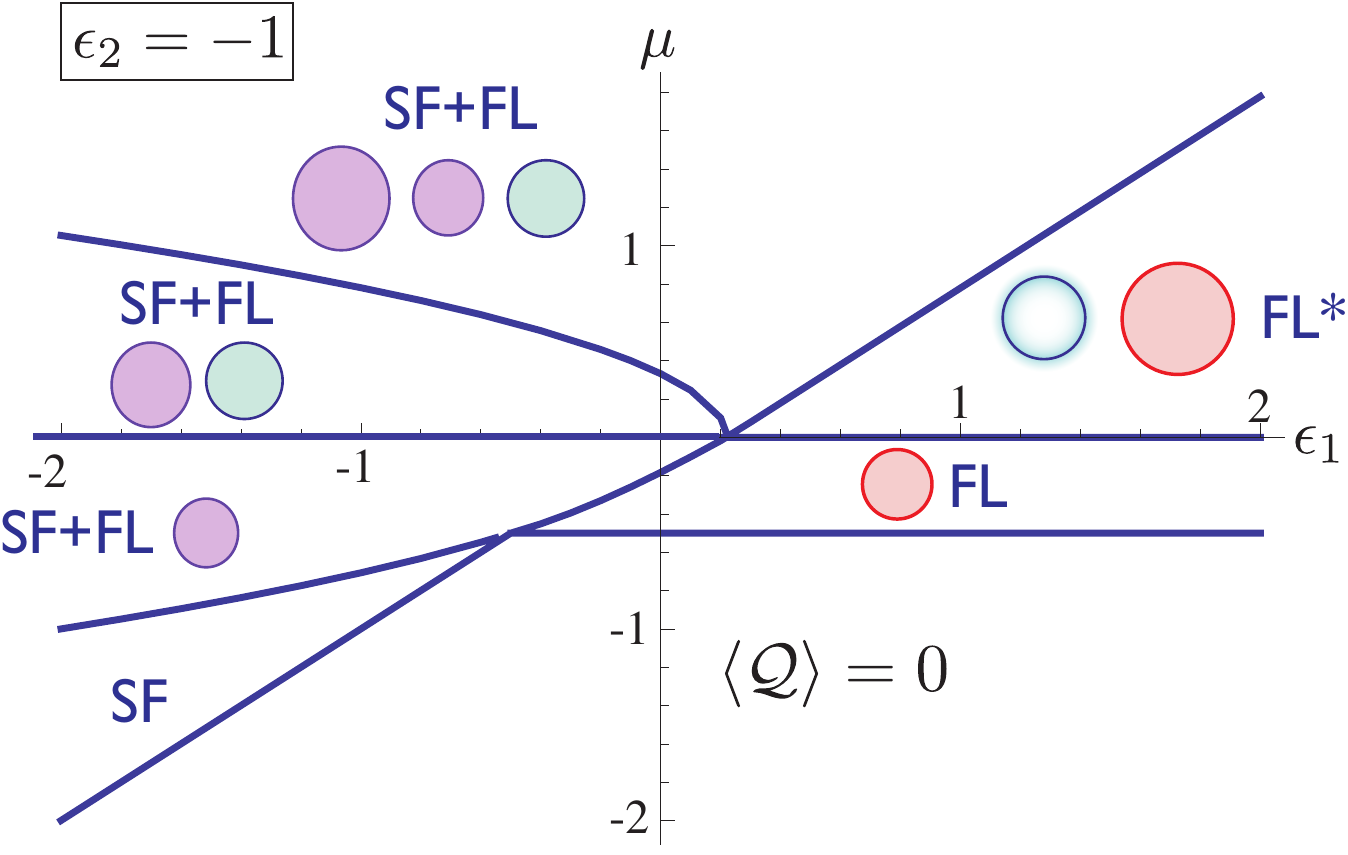}
\end{center}
\caption{As in Fig.~\ref{pd_abjm1}, with $g_1 = 0$ and $g_2 = 1$, but with a different value of $\epsilon_2$. In this and the following figures
in Section~\ref{sec:ABJM},
we have $\langle b_+ \rangle = \langle b_- \rangle \neq 0$ in phase with SF or FL labels, unless otherwise noted.}
\label{pd_abjm2}
\end{figure}
\begin{figure}[h]
\begin{center}
\includegraphics[width=5in]{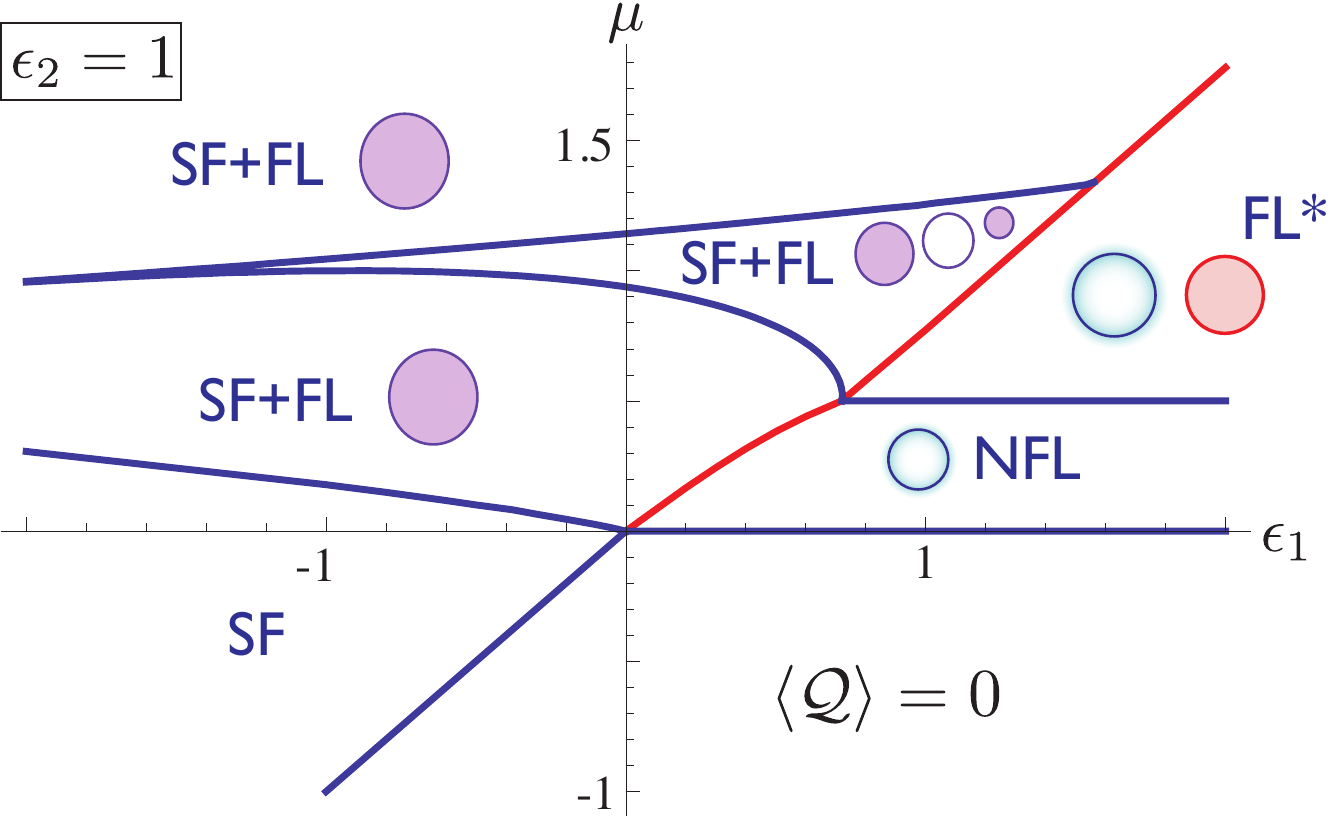}
\end{center}
\caption{As in Fig.~\ref{pd_abjm1}, but with $g_1 = 1$ and $g_2 = 1$. Furthermore, we have set $v=0.85$. The red line represents a first-order transition. The onset of SF order from the FL* or NFL phases is required to be first order because of reasons discussed near 
Eq.~(\ref{4log}). The SF+FL phase with 3 Fermi surfaces has a non-monotonic fermionic dispersion, and so has 3 Fermi surfaces; one of these
carries $\mathcal{Q}=-1$ ({\em i.e.\/} it is ``hole''-like), and is indicated by the unfilled circle.}
\label{pd_abjm5}
\end{figure}
\begin{figure}[h]
\begin{center}
\includegraphics[width=5in]{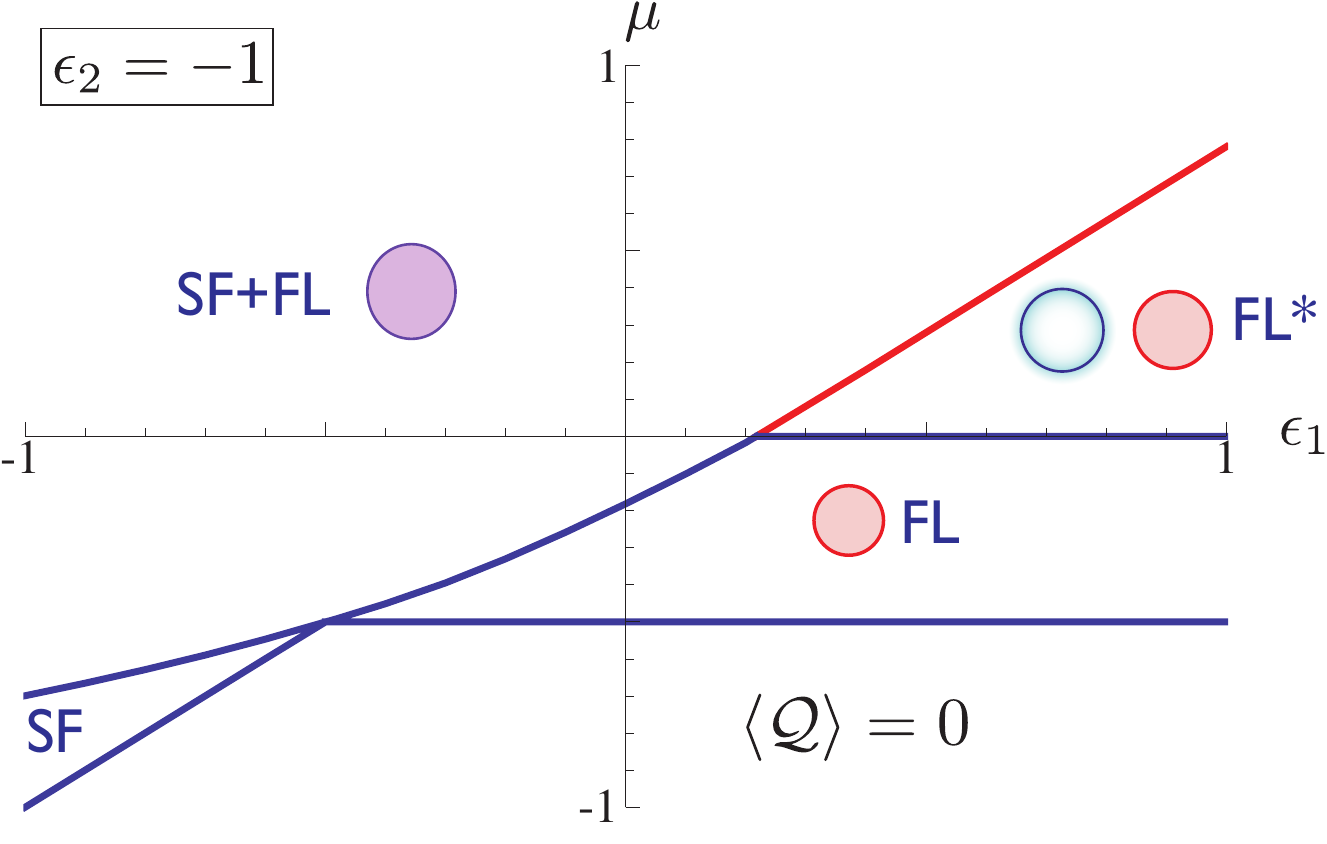}
\end{center}
\caption{As in Fig.~\ref{pd_abjm5}, with $g_1 = 1$, $g_2 = 1$ and $v=0.85$, but with a different value of $\epsilon_2$.}
\label{pd_abjm6}
\end{figure}

It is useful to now define new canonical Fermi operators 
\bea
f_1 &=& \frac{1}{\sqrt{|b_+|^2 + |b_-|^2}} \left( f_{+}b_{-} + f_{-}b_{+} \right) \nn
f_2 &=& \frac{1}{\sqrt{|b_+|^2 + |b_-|^2}} \left(  - f_{+}b_{+}^\ast + f_{-}b_{-}^\ast \right) . \label{f1f2}
\eea
Note that the $c$ fermion couples only to $f_1$, and $f_1 f_2 = f_+ f_-$, this enables diagonalization of the mean-field
fermion Hamiltonian.
The mean-field energy density at $T=0$ is
\bea
E(b_{+},b_{-}) &=& (\epsilon_1 - \mu)  \left( |b_{+}|^{2} + |b_{-}|^{2} \right) + \frac{u}{2}\left( |b_{+}|^{2} + |b_{-}|^{2} \right)^{2}
+ v \, |b_{+}|^{2} |b_{-}|^{2} \nn &+& \int \frac{d^{2}k}{4 \pi^{2}} \left[\sum_{j=1}^{3} \Bigl\{ \varepsilon_{j} (k) \theta(-\varepsilon_{j}(k)) \Bigr\}+ \frac{k^{2}}{2 m_{f}} - \mu + \frac{g_{1}^{2} |b_+|^2 |b_-|^2 m_{f}}{k^{2} + \Lambda^{2}} \right],
\label{ee}
\eea
where $\theta$ is the unit step function, and $\varepsilon_j (k)$ are the 3 eigenvalues of the matrix
\beq
M(k) = \left(
\begin{array}{ccc}
 \dfrac{k^2}{2 m_c} + \epsilon_2 - 2 \mu  & - g_2 \sqrt{|b_+|^2 + |b_-|^2} & 0 \\
 \noalign{\medskip}
- g_2 \sqrt{|b_+|^2 + |b_-|^2} &  \dfrac{k^2}{2 m_f}  -  \mu & - g_1 b_+ b_- \\
 \noalign{\medskip}
0 & - g_1 b_+^\ast b_-^\ast & - \dfrac{k^2}{2 m_f}  +   \mu
\end{array}
\right).
\eeq
We have renormalized the coupling $v$ by
\beq
v \rightarrow v +  \int \frac{d^{2}k}{4 \pi^{2}} \frac{g_{1}^{2} m_{f}}{k^{2} + \Lambda^{2}},
\eeq
where $\Lambda$ is a renormalization scale, so that the momentum integral in \eqnref{ee} is ultraviolet convergent.
To determine the phase diagram we now have to minimize the function in \eqnref{ee} with respect to the
complex numbers $b_\pm$, while fixing the density by
\beq
- \frac{\partial E}{\partial \mu} = \left\langle \mathcal{Q} \right\rangle .
\eeq
We also have to maintain global neutrality of the U(1) gauge charge, as in Section~\ref{sec:ffl}, and so we have the constraint
\beq
\langle b_+^\dagger b_+ \rangle  + \langle f_+^\dagger f_+ \rangle = \langle b_-^\dagger b_- \rangle  + \langle f_-^\dagger f_- \rangle.
\label{bfconst}
\eeq

The results of such an energy minimization under the constraint in Eq.~(\ref{bfconst}) are shown in the phase diagrams of Figs.~\ref{pd_abjm0}-\ref{pd_abjm6}.
We choose parameters $m_f=m_b=m_c/2=u=1$ and others as specified in the figures.

The phases are distinguished by the nature of the $b_\pm$ condensates, and the configurations of the Fermi surfaces:
\begin{enumerate}
\item $\langle b_+ \rangle = \langle b_- \rangle = 0$. The bosons are gapped and we need only pay attention to the Fermi surfaces.
Because the global U(1) symmetry associated with the charge $\mathcal{Q}$ in Eq.~\ref{Qabjm} is realized, there is a Luttinger relation constraining the volumes of the Fermi surfaces:
\beq
2 V_f + 2 V_c = \left\langle \mathcal{Q} \right\rangle;
\eeq
the prefactor in front of $V_f$ arises from the sum over $f_+$ and $f_-$ Fermi surfaces, while the prefactor of $V_c$ is from the 
$c$ charge in Eq.~(\ref{Qabjm}). The phases here are further subclassified by the configurations of the Fermi surfaces, as shown in 
Fig.~\ref{pd_abjm0}:
\begin{enumerate}
\item \underline{FL:}
If $V_f=0$, then there are no Fermi surfaces with gauge charges, the U(1) gauge field
is confining (the U(1) is presumed to be embedded in a compact gauge group), and we obtain an FL phase with only a $c$
Fermi surface. 
\item \underline{FL*:} Now both $V_f$ and $V_c$ are non-zero. The $f_\pm$ fermions are coupled to the U(1) gauge field,
which is now in a deconfined phase. This phase is similar to the FL* phase of Section~\ref{sec:ffl}.
\item \underline{NFL:} This non-Fermi liquid phase has $V_c=0$, the U(1) gauge force is deconfined, and the phase is similar to that in Section~\ref{sec:doublon}.
\end{enumerate}
As discussed in Section~\ref{sec:doublon}, we expect the NFL and FL* phases to be ultimately unstable to
fermion pairing induced by exchange of gauge bosons. However, this is a fluctuation correction to our mean field theory,
and its importance in the large $N$ limit of the gauge theory remains to be studied.
\item 
$\langle b_+ \rangle = \langle b_- \rangle \neq 0$. Both bosons condense and gap out the U(1) gauge field, leading to a confining
phase \cite{peskin}.
Notice that the product $\langle b_+ \rangle \langle b_- \rangle $ is a gauge-invariant condensate which carries
the global $\mathcal{Q}$ charge: consequently the global U(1) symmetry is also broken and such phases are superfluids. They
are expected to correspond to the superfluids found in holographic studies \cite{denef0,gubser,hhh,sonner,gubserpufurocha,Donos:2010ax,gauntlett}.
Note that such superfluids correspond to a gauge-invariant condensate with charge $\mathcal{Q}=2$. 
This is confirmed by the presence of half-vortices \cite{wenlee,balents1,balents2}: such a vortex has $b_+ \sim e^{i \theta}$
(where $\theta = \tan^{-1}(y/x)$ is the azimuthal angle), $b_- \sim 1$, U(1) gauge flux $\int d^2 r {\bm \nabla} \times {\bm A} = \pi$, and also $\pi$ magnetic flux dual to the $\mathcal{Q}$ charge.
As in Section~\ref{sec:bf}, there is no constraint on the Fermi surface volumes
when superfluidity is present.
If no Fermi surfaces are present, we obtain
a SF phase. One or more Fermi surfaces can also be present: in general these will be Fermi surfaces of hybridized $c$ and $f_\pm$ fermions, 
and will carry Fermi liquid quasiparticles, and so such phases are SF+FL.
\item 
$\langle b_+ \rangle \neq 0$ or $\langle b_- \rangle \neq 0$, but not both.
Now the U(1) gauge field is gapped and this is a confining phase\cite{peskin}, 
but there is no gauge-invariant observable which carries the global $\mathcal{Q}$ charge.
So there is no SF order. Stability requires that the total gauge charge be zero (as in Eq.~(\ref{bfconst})), and so there is
a compensating Fermi surface of $f_-$ fermions to achieve this. So this phase is a
FL, with one or more Fermi surfaces of $f_\pm$, $c$, or their hybridized combinations.
This phase breaks the $Z_2$ symmetry mentioned below Eq.~(\ref{l1}). Such phases appear only at very small
values of $\mu$ and $\epsilon_2$ in our mean-field phase diagrams, which we discuss in Appendix~\ref{app:gauge}.
\end{enumerate}

\section{Theory similar to $\mathcal{N}=4$ Super Yang Mills in $d=3$}
\label{sec:SYM}

The SU($N$) Yang-Mills gauge theory in $d=3$ and $\mathcal{N}=4$ supersymmetry (SYM4) is 
the simplest and best-studied case of gauge-gravity duality. It should therefore pay to also exploit
it to understand gravity duals of systems with Fermi surfaces.

As in the $d=2$ case considered in \secref{sec:ABJM}, the gauge theory has supersymmetry and global symmetries which are broken
by the application of chemical potential. Our strategy will be to write down the simplest model
with a similar particle content which is consistent with the residual symmetries.

We review the particle content and symmetries of SYM4 in Appendix~\ref{app:SYM} from a condensed-matter perspective.
The theory has fermions $\lambda_{i\alpha}^{a}$ with adjoint color $a$, SU(4) flavor $i$, and Weyl spinor $\alpha$ indices.
Adding the three possible chemical potentials reduces the SU(4) symmetry to U(1)$\times$U(1)$\times$U(1). 
For now we consider the case where only a single chemical potential, say $\mu_{1}$, is non-zero; other cases will be
discussed in Section~\ref{sec:chem}. For the single chemical potential case, 
the flavor symmetry is reduced to U(1)$\times$SO(4). We simplify the theory further by dropping the SO(4) symmetry, and hence
the flavor index $i$, as in \secref{sec:ABJM}. As discussed in Appendix~\ref{app:SYM}, the fermions form bosonic pairs which are antisymmetric
in color, flavor, and spin, and this is consistent with overall fermionic antisymmetry.
We want to retain antisymmetry in color, we have already dropped flavor, and so let us also drop the Weyl spin index.
This is natural from the absence of relativistic invariance in the presence of the chemical potential.
So we are led to consider a theory with non-relativistic fermions $\lambda^{a}$ which has a global U(1) symmetry,
under which $\lambda^{a} \rightarrow e^{i \theta} \lambda^{a}$, and an adjoint color index $a=1 \ldots N^{2}-1$.

We can apply a similar reasoning to the scalar sector. The SYM theory of Appendix~\ref{app:SYM} has complex scalars 
$\Phi^{a}_{p}$, where $p=1,2,3$ is an index specifying transformations under SU(4). With our choice of chemical potential, $\mu_1 \neq 0$,
the $\Phi^{a}_1$ scalar is preferred: we will only work with this scalar, dropping the $\Phi^{a}_{2,3}$ scalars whose spectrum remains
relativistic. So we drop the SU(4) $p$ index, work with a non-relativistic
kinetic energy, but retain a global U(1) under which $\Phi^{a} \rightarrow e^{2i \theta} \Phi^{a}$: this transformation reflects the fact
that the scalar couples to fermion pairs in \eqnref{LY}.

Finally, we retain the SU($N$) gauge field, $(A_{\tau}^{a}, {\bf A}^{a})$, of SYM4 with no changes. However, our mean-field theory
below will not explicitly include gauge fluctuations.

We can now write down the simplest Lagrangian containing these fields and consistent with symmetries.
We generalize the temporal derivative $\partial_{\tau}$ to  the covariant derivative $D_{\tau}$, and the 
spatial gradient ${\bm \nabla}$ to the covariant derivative ${\bm D}$: see Appendix~\ref{app:SYM} for explicit
expressions with all indices. Then we have
\bea \mathcal{L}_{2} &=&  \mathcal{L}_{\Phi} + \mathcal{L}_\lambda    \nn
\mathcal{L}_\Phi &=& \Phi^{\dagger} \left( D_{\tau} - 2 \mu + \epsilon_{1}
 -  \frac{{\bm D}^{2}}{2m_1} \right)  \Phi + 
u  \left( \Phi^{a\dagger} \Phi^{a} \right)^2 \nn
\mathcal{L}_\lambda &=&  \lambda^{\dagger} \left(D_{\tau} - \mu -  \frac{{\bm D}^{2}}{2m_2}  \right)  \lambda
+  g_1 \left( f_{abc} \Phi^{a\dagger} \lambda^b \lambda^{c} + {\rm c.c.} \right) \label{L0}
\eea
where the chemical potential $\mu = \mu_{1}/2$, with $\mu_1$ as defined in Appendix~\ref{app:SYM}, $f_{abc}$ are the structure constants of SU($N$). For $N=2$, $f_{abc} = \epsilon_{abc}$.
We have inserted a quartic scalar coupling $u$ to prevent runaways in the scalar, and stabilize the theory.
The chemical potential $\mu$ couples to the global U(1) charge, and $\epsilon_{1}$ is a parameter
we will use to tune between possible phases. The coupling $g_{1}$ mirrors the Yukawa coupling
of SYM4 in \eqnref{LY}.

As we noted above, we do not account for SU($N$) gauge field fluctuations in the mean-field
analysis below. It is therefore useful to include additional effective interaction terms in our theory which account for the gauge forces, and are easier
to include in mean-field theory. As we discuss in Appendix~\ref{app:SYM}, one effect of the gauge forces is to bind fermions into pairs
which are antisymmetric in color: this pair binding effect is already included via the $g_{1}$ coupling in $\mathcal{L}_{\lambda}$.
For the comparison with dual gravity theories, we would also like a gauge-invariant fermion, and so let us include interactions
which favor singlet bound states of 3 fermions. In terms of $\Phi^{a}$, such a term can be written as the attractive interaction
$- \Phi^{a\dagger} \lambda^{a\dagger} \lambda^{b} \Phi^{b}$, which is analogous to the boson-fermion interaction in \eqnref{Lbf}. 
Just as in \eqnref{Lbfc}, we can decouple this interaction by introducing a color-singlet fermionic field $c$.
As in \secref{sec:ABJM}, there can be numerous such bound states of the gauge-charged bosons and fermions \cite{sean2,sean3},
but, for simplicity, we will only include a single
gauge-invariant and canonical fermion.
Our theory becomes
\bea \mathcal{L}_{3} &=&  \mathcal{L}_{2} + \mathcal{L}_c    \nn
\mathcal{L}_c &=&  c^{\dagger} \left(\partial_{\tau} - 3 \mu + \epsilon_{2} -  \frac{{\bm \nabla}^{2}}{2m_3}    \right) c 
+  g_2 \left(  c^{\dagger} \lambda^a \Phi^{a} + {\rm c.c.} \right) \label{L}
\eea
where $\epsilon_{2}$ is another tuning parameter.
For our final theory $\mathcal{L}_{3}$, the global conserved U(1) charge is
\beq
\mathcal{Q} = \lambda^{a\dagger} \lambda^{a} + 2 \Phi^{a \dagger} \Phi^{a} + 3 c^{\dagger} c. \label{qsym}
\eeq

The present theory $\mathcal{L}_3$ is similar to the ABJM-inspired 
theory $\mathcal{L}_1$ in Eq.~(\ref{l1}), but differs from it in
a crucial respect: the $g_1$ fermion pairing term in Eq.~(\ref{L0}) couples to a single boson $\Phi^a$, while
the pairing term in Eq.~(\ref{l1}) coupled to a boson pair $b_+ b_-$. This has the important consequence that the present
model $\mathcal{L}_{3}$ has a BCS-like instability of $\lambda^a$ Fermi surfaces to the onset of SF order even in mean-field 
theory, while the ABJM-inspired model $\mathcal{L}_{1}$ does not. Specifically, let us assume we are in a phase in which $\lambda^a$ Fermi surfaces are present, while $\langle \Phi^a \rangle \neq 0$. Now integrate out the $\lambda^a$ fermions from $\mathcal{L}_3$, and compute the free energy
to order $g_1^2$. This leads to the familiar BCS log divergence, and a contribution to the ground state energy of the form
\beq
E \sim g_1^2 |\Phi^a |^2 \log ( |\Phi^a |^2) + \ldots \label{bcslog}
\eeq
while other terms are smooth functions of $|\Phi^a|^2$. It is not possible for such an expression to have a stable minimum at $\Phi^a = 0$. The conclusion is that any phase with $\lambda^a$ Fermi surfaces is necessary unstable to the appearance of SF order.
Hence there can be no FL* or NFL phases in the mean-field phase diagram of $\mathcal{L}_3$, and we will see that is indeed the
case in our analysis below. 

In contrast for the ABJM-inspired model $\mathcal{L}_1$ in Eq.~(\ref{l1}) the corresponding contribution to the energy has the
form
\beq
E \sim g_1^2 |b_+|^2 |b_- |^2 \log \left[ |b_+|^2 |b_-|^2 \right] + \ldots \label{4log}
\eeq
and this can have a stable minimum at $b_\pm =0$. This fact accounts for the presence of the FL* and NFL phases
in Figs.~\ref{pd_abjm0}-\ref{pd_abjm6}. This logarithm also requires the transition of the onset of SF order from
the FL* and NFL phases to be first order, as is the case in Figs.~\ref{pd_abjm5} and~\ref{pd_abjm6}.

\subsection{Phase diagram}
\label{sec:pdsym}

We proceed as in Section~\ref{sec:pdabjm}, and compute the mean-field phase diagram of
the SYM-inspired theory $\mathcal{L}_3$ in Eq.~(\ref{L}).

\begin{figure}[t]
\begin{center}
\includegraphics[width=5in]{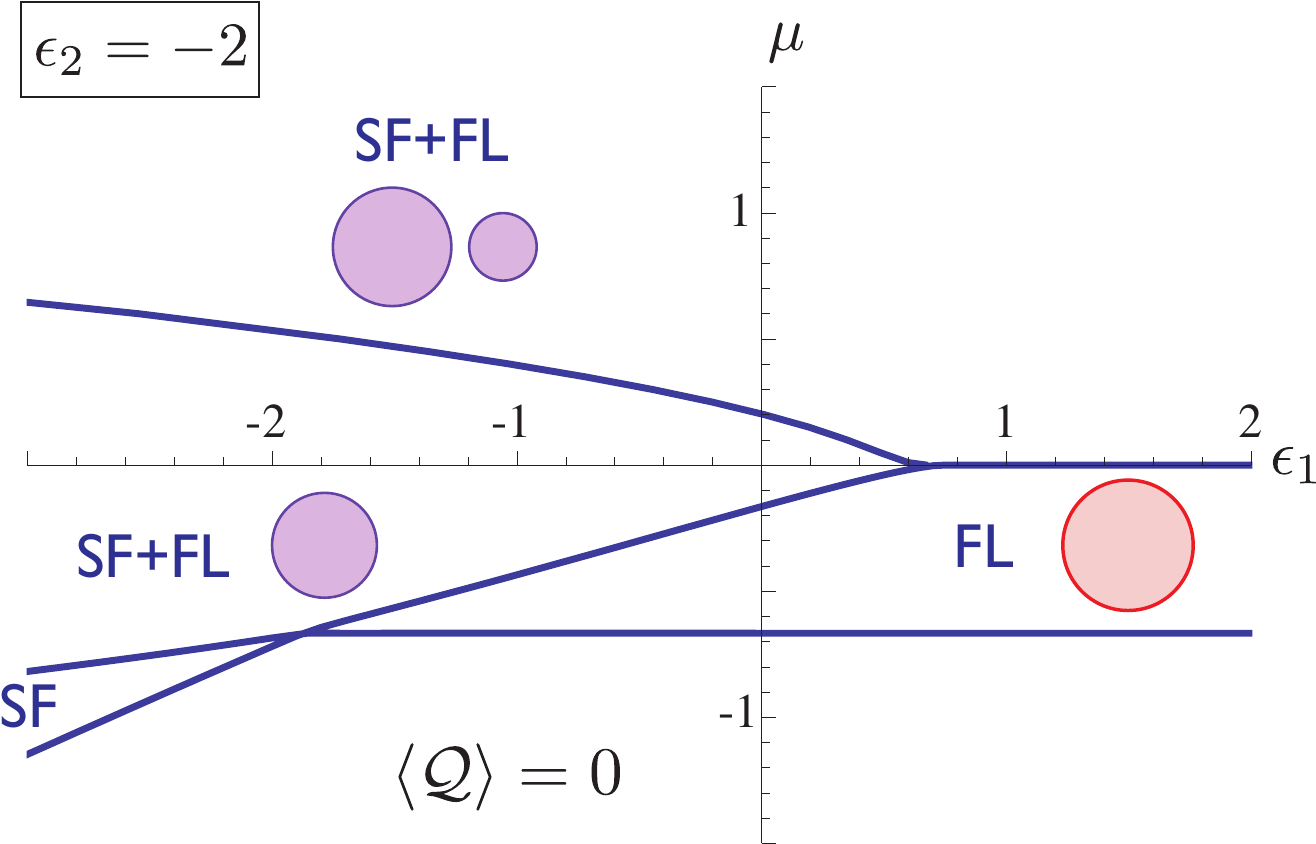}
\end{center}
\caption{Mean field phase diagram of the SYM-inspired theory $\mathcal{L}_3$ in Eq.~(\ref{L}). 
This theory is similar to the SYM model of Appendix~\ref{app:SYM} with only the chemical potential $\mu_1 \neq 0$.
All the phases labelled SF have $\langle \Phi^a \rangle \neq 0$, while the remainder have $\langle \Phi^a \rangle = 0$.
We use
the same color conventions as in Fig.~\ref{pd_abjm1}, with the $f$ fermions replaced by the $\lambda$ fermions: 
the purple Fermi surfaces contain hybridized combinations of 
the $\lambda^a$ and $c$ fermions. Unlike Figs.~\ref{pd_abjm1} and~\ref{pd_abjm2}, notice there are 
now no NFL or FL* phases; this is explained by Eq.~(\ref{bcslog}).}
\label{pd_sym1}
\end{figure}

\begin{figure}[t]
\begin{center}
\includegraphics[width=5in]{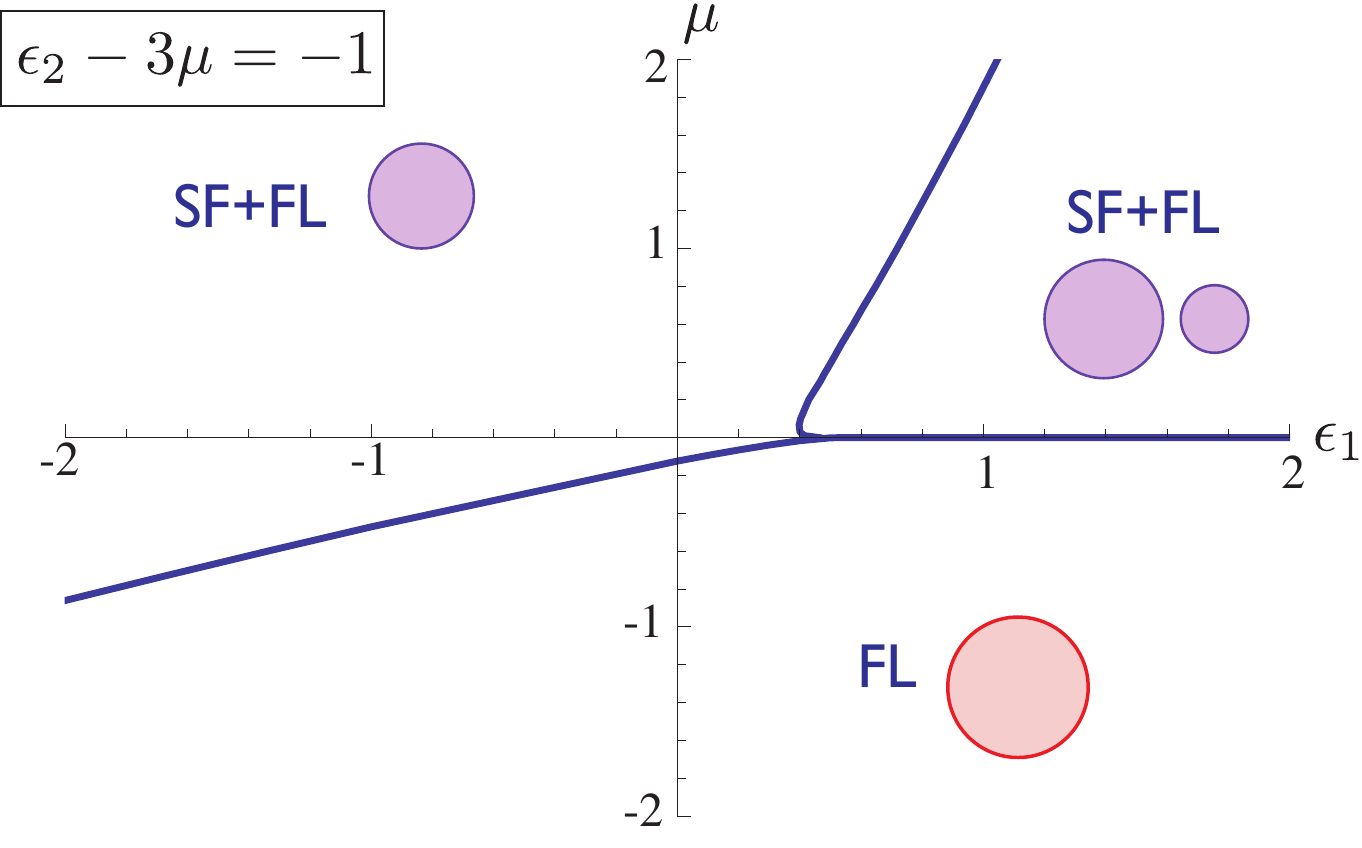}
\end{center}
\caption{As in Fig.~\ref{pd_sym1}, but with a different value of $\epsilon_2$.}
\label{pd_sym2}
\end{figure}

We will consider the case with a SU(2) gauge field, and so $N=2$, $a,b,c=1 \ldots 3$, and $f_{abc}=\epsilon_{abc}$ the anti-symmetric tensor.
The mean field Hamiltonian for the fermions follows from setting the $\Phi^a$ to constants. Without loss of generality, we can perform
a global SU(2) rotation to replace 
$\Phi^a$ by the vector $(0,\Delta_1,\Delta_2)$, where $\Delta_1$ is real and $\Delta_2$ is complex. 
The value of $\Phi^a$ is also restricted by the requirement of gauge charge neutrality, as we will discuss below.
With this choice the $g_1$- and $g_2$-terms read:
\bea
2 g_1 \left[ \l^1 (\Delta_2^* \l^2 - \Delta_1 \l^3) + (\Delta_2 \l^{2 \dag} - \Delta_1 \l^{3 \dag}) \l^{1 \dag} \right] \nn
+ g_2 \left[ c^{\dag}  (\Delta_1 \l^2 + \Delta_2 \l^3) + (\Delta_1 \l^{2 \dag} + \Delta_2^* \l^{3 \dag}) c \right] .
\eea
This suggests we should introduce the fermions
\bea
f_+ &=& (\Delta_1 \l^2 + \Delta_2 \l^3)/\Delta \nn
f_- &=& (\Delta_2^* \l^2 - \Delta_1 \l^3)/\Delta ,
\eea
where we defined $\Delta \equiv \sqrt{\Delta_1^2 + |\Delta_2|^2}$. One readily checks that we now have
\bea
f_+^{\dag} f_+ +f_-^{\dag} f_- = \l^{2 \dag} \l^2 + \l^{3 \dag} \l^3.
\eea
Consequently, the mean field Hamiltonian can be written as
\bea
H_{\text{mf}} &=& \int \frac{d^3 k}{(2 \pi)^3} \left[ \xi^{\l}_k (\l^{1 \dag} \l^1 + f_+^{\dag} f_+ +f_-^{\dag} f_-) + \xi^c_k c^{\dag} c  + 2 g_1 \Delta ( \l^1 f_- + f_-^{\dag} \l^{1 \dag} )   \right . \nn
& & \left.+ g_2 \Delta (c^{\dag} f_+ + f_+^{\dag} c ) \right] -(2 \mu - \epsilon_1) \Delta^2 + u \Delta^4
\eea
where we defined
\bea
 \xi^{\l}_k &=& \frac{k^2}{2m_2}-\mu ,\nn
 \xi^{c}_k &=& \frac{k^2}{2m_3}-3\mu+ \epsilon_2.
\eea
Introducing the mixed fermions $F_{\pm}, \Psi_{\pm}$ the Hamiltonian takes on the diagonal form:
\bea
H_{\text{mf}} &=& \int \frac{d^3 k}{(2 \pi)^3} \left[ \xi^{F_+}_k F_+^{\dag} F_+ + \xi^{F_-}_k F_-^{\dag} F_- + \xi^{\Psi_+}_k \Psi_+^{\dag} \Psi_+ + \xi^{\Psi_-}_k \Psi_-^{\dag} \Psi_- + \xi^{\l}_k  \right] \nn
& & -(2 \mu - \epsilon_1) \Delta^2 + u \Delta^4,
\eea
with
\bea
 \xi^{F_{\pm}}_k &=& \frac{1}{2} (\xi^c_k + \xi^{\l}_k) \pm \frac{1}{2} \sqrt{ (\xi^c_k - \xi^{\l}_k)^2 + 4 g_2^2 \Delta^2} ,\nn
 \xi^{\Psi_{\pm}}_k &=& \pm \sqrt{(\xi^{\l}_k)^2 + g_1^2 \Delta^2 }.
\eea
We can now write down the free energy:
\bea\label{FE}
E (\Delta) =  -(2 \mu - \epsilon_1) \Delta^2 + u \Delta^4 + \sum_{x \in \{F_{\pm},\Psi_{\pm} \}} R^x (\Delta) + \int \frac{d^3 k}{(2 \pi)^3} \xi^{\l}_k,
\eea
with
\bea
 R^x (\Delta)  = -T \int \frac{d^3 k}{(2 \pi)^3} \ln (1 + e^{-  \xi^x_k /T}).
\eea
For $T=0$ this reduces to
\bea
 R_0^x (\Delta)  = \int \frac{d^3 k}{(2 \pi)^3} \theta(-\xi^x_k) \xi_k^x.
\eea
It follows that $ R_0^{\Psi_+} (\Delta)=0$, since $\xi_k^{\Psi_+}>0$ for all $k$. Furthermore, we find that $ R_0^{\Psi_-} (\Delta)$ diverges for large energies:
\bea
 R_0^{\Psi_-} (\Delta)  &=& - \int \frac{d^3 k}{(2 \pi)^3}  \sqrt{(\xi^{\l}_k)^2 + g_1^2 \Delta^2 } \nn
&\approx& - \int \frac{d^3 k}{(2 \pi)^3} \frac{k^2}{2 m_2} (1 +\frac{1}{2}\frac{g_1^2 \Delta^2}{(k^2/(2m_2))^2} + \dots ).
\eea
The first term is cancelled by the last term in (\ref{FE}), but the second term is also divergent. This is remedied by adding and subtracting the terms
\bea
 \int \frac{d^3 k}{(2 \pi)^3}  \frac{m_2 g_1^2 \Delta^2}{k^2} - \int\frac{d^3 k}{(2 \pi)^3} \frac{m_2 g_1^2 \Delta^2}{k^2}.
\eea
The first term cancels the divergence and the second term is absorbed by renormalizing the detuning $\epsilon_1$:
\bea
\epsilon_1 \to \epsilon_1  -  \int\frac{d^3 k}{(2 \pi)^3} \frac{m_2 g_1^2 \Delta^2}{k^2}.
\eea
Putting everything together we obtain
\bea\label{FEa}
E (\Delta) &=&  -(2 \mu - \epsilon_1) \Delta^2 + u \Delta^4  - \int \frac{d^3 k}{(2 \pi)^3}  \sqrt{(\xi^{\l}_k)^2 + g_1^2 \Delta^2 }  \nn
& & +  \int \frac{d^3 k}{(2 \pi)^3}  \frac{m_2 g_1^2 \Delta^2}{k^2} + \int \frac{d^3 k}{(2 \pi)^3} \xi^{\l}_k  + \int \frac{d^3 k}{(2 \pi)^3} \theta(-\xi^{F_+}_k) \xi_k^{F_+} \nn
& & + \int \frac{d^3 k}{(2 \pi)^3} \theta(-\xi^{F_-}_k) \xi_k^{F_-} .
\eea
Now we minimize $E (\Delta)$ as a function of $\Delta$, for $u=1$, $m_1=m_2=m_3/2=3.13$, 
and so obtain the phase diagram shown in Figs.~\ref{pd_sym1}
and~\ref{pd_sym2}.

The description of the phases in the phase diagram closely parallels the analysis of Section~\ref{sec:pdabjm}. 
Now all phases with $\langle \Phi^a \rangle \neq 0$ also have SF order, because the gauge invariant combination
$\langle \Phi^a \Phi^a \rangle \approx \langle \Phi^a \rangle \langle \Phi^a \rangle$ carries the global charge $\mathcal{Q}$ of Eq.~(\ref{qsym}).
Thus the superfluid condensate has charge $\mathcal{Q}=4$. 
The condensate cannot carry a net SU(2) gauge charge \cite{rmp}, and so we should have 
\beq
\epsilon_{abc} \langle \Phi^{b \dagger} \Phi^c \rangle = 0, 
\label{neutral}
\eeq
which is realized in mean field theory by a real $\Delta_1$. Actually the restriction is on total gauge charge neutrality,
including the contributions of the fermions. However, 
time-reversal symmetry and  a U(1) particle-number transformation ensure that Eq.~(\ref{neutral}) 
also implies neutrality of the fermion contribution.
Under time-reversal,
\bea
\lambda^a (\tau) \rightarrow \lambda^{a \dagger} (-\tau) \quad &,& \quad
\lambda^{a \dagger} (\tau)  \rightarrow  -\lambda^a (-\tau) \nn
\Phi^a (\tau)  \rightarrow  -\Phi^{a \dagger} (-\tau) \quad &,& \quad
\Phi^{a \dagger} (\tau)  \rightarrow  -\Phi^a (-\tau) \nn
A_{\tau}^a (\tau)  \rightarrow  -A_{\tau}^a (-\tau) \quad &,& \quad
{\bf A}^{a}(\tau)  \rightarrow  {\bf A}^{a}(-\tau),
\eea
and so the gauge charge changes sign $Q^a \rightarrow -Q^a$.
Combining time reversal with a U(1) transformation associated with $\mathcal{Q}$,
$\lambda^a \rightarrow i \lambda^a$, $\Phi^a \rightarrow - \Phi^a$
we have $\Phi^a \rightarrow \Phi^{a \dagger}$ and $Q^a \rightarrow -Q^a$. 
Hence Eq.~(\ref{neutral})
guarantees that $\langle Q^a \rangle = 0$.

Let us also examine the structure of the fluctuations of $A_\mu^a$, the SU(2) gauge field, in the phases with $\langle \Phi^a \rangle \neq 0$.
The gluon mass terms generated by the boson condensate are
\beq
\left| \epsilon_{abc} A_\mu^b \Phi^c \right|^2 = | A_\mu^2 \Delta_2 - A_\mu^3 \Delta_1 |^2 + \left(A_\mu^1 \right)^2 \left| 
\Delta \right|^2 
\eeq
So the field $A_\mu^1$ is massive, and we will drop it from now. Diagonalizing the quadratic form, we see that the linear combination
\beq 
A_\mu^+ = (\Delta_1 A_\mu^2 + \Delta_2 A_\mu^3)/\Delta
\eeq
remains gapless if $\mbox{Im} \left( \Delta_1 \Delta_2^\ast \right) = 0$, which is indeed the case from Eq.~(\ref{neutral}); the orthogonal combination $A_\mu^-$ is gapped and will also be dropped.
The coupling of the fermions to the temporal component of the $A_\mu^+$ gauge field is
\beq
\epsilon_{abc} \lambda^{a\dagger} A_\tau^b \lambda^c = A_\tau^+ ( f_-^\dagger \lambda^1 - \lambda^{1 \dagger} f_- )
\eeq
Notice that the gapless gauge field does not couple to the fermions $c$ and $f_+$ which form the Fermi surfaces.
Therefore this gapless component will not be damped by Fermi surface excitations, and so will eventually confine.
So all the gauge field components are gapped in the SF phase with $\langle \Phi^a \rangle \neq 0$. Thus we expect that all the
SF phases are smoothly connected \cite{smooth1,smooth2} to a description of the superfluidity in terms of the condensation of the gauge neutral scalar
$\Phi^a \Phi^a$ carrying charge $\mathcal{Q}=4$. 
These SF phases are expected to be related to the holographic superfluids \cite{gubser,hhh,gubserpufurocha,pufu,leigh,aprile}.

The $\mathcal{Q}=4$ charge of the SF condensate is confirmed to the presence of half-vortices, as in Section~\ref{sec:ABJM}.
Here, we can write the field configuration of around a half-vortex as $\Phi^a = e^{i \theta/2} R^a$, where $\theta = \tan^{-1}(y/x)$ is the azimuthal angle, and $R^a$ is a real 3-component vector which traces a curve on $S^2$ 
from the north pole to the south pole as the vortex is encircled.

We turn to phases with $\langle \Phi^a \rangle = 0$. These can only be stable if they
don't have $\lambda^a$ Fermi surfaces, as noted with Eq.~(\ref{bcslog}). Such phases can therefore only have $c$ Fermi surfaces, and so must
be FL states with the gauge theory in a confining phase; from Eq.~(\ref{qsym}), they obey the Luttinger constraint
\beq
3 V_c = \left\langle \mathcal{Q} \right\rangle.
\eeq

Note that all the phases of $\mathcal{L}_3$ in Figs.~\ref{pd_sym1} and~\ref{pd_sym2} have the gauge field in Higgs/confining mode,
and there are no deconfined phases. This feature differs from the ABJM-like model in Section~\ref{sec:SYM},
and for the SYM-like model with distinct chemical potential assignments to be considered in the following subsection. 

\subsection{Other chemical potential choices}
\label{sec:chem}

So far, we have discussed the case inspired by the choice of the chemical potentials of the SYM theory 
$\mu_1 \neq 0$, $\mu_2=\mu_3=0$, in the
notation of Appendix~\ref{app:SYM}.  The case $\mu_1=\mu_2=\mu_3 \neq 0$, which leads to the 
extensively studied Reissner-Nordstrom black holes \cite{nernst,sslee0,hong0,zaanen1,hong1,denef}, is connected to models
very similar to those already considered.
In this section we consider models inspired by the case 
$\mu_1 = \mu_2 \neq 0$, $\mu_3 =0$, which leads to some qualitatively different physics. This case is related to the model
studied by Gubser and Rocha \cite{gubserrocha}.

\begin{figure}[h]
\begin{center}
\includegraphics[width=5in]{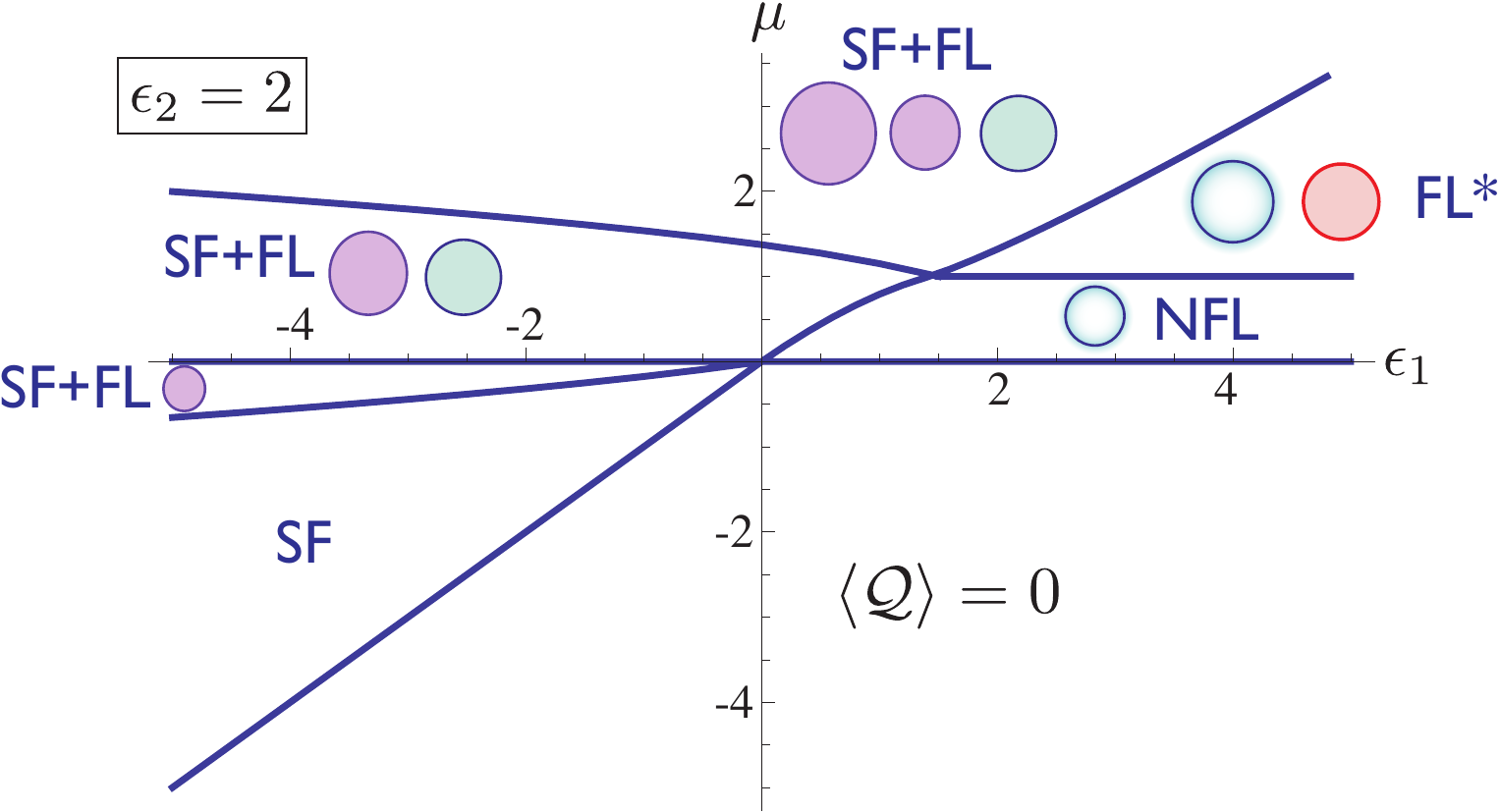}
\end{center}
\caption{Mean field phase diagram of the SYM-inspired theory $\widetilde{\mathcal{L}}_3$ in Eq.~(\ref{tildeL}). 
This theory is similar to the SYM model of Appendix~\ref{app:SYM} with 2 chemical potentials non=zero: $\mu_1 = \mu_2 \neq 0$.
All the phases labelled SF have $\langle \Phi^a \rangle \neq 0$, while the remainder have $\langle \Phi^a \rangle = 0$.
The coloring conventions are as in Fig.~\ref{pd_sym1} and~\ref{pd_abjm1}, and the blue Fermi surfaces represent $\lambda^a$
fermions. Unlike the SYM-like model considered previously
in Fig.~\ref{pd_sym1} and~\ref{pd_sym2}, now there are deconfined phases with Fermi surfaces of $\lambda^a$ fermions
coupled to gapless gauge field: these are the FL* and NFL phases.
}
\label{pd_sym3}
\end{figure}
\begin{figure}[h]
\begin{center}
\includegraphics[width=5in]{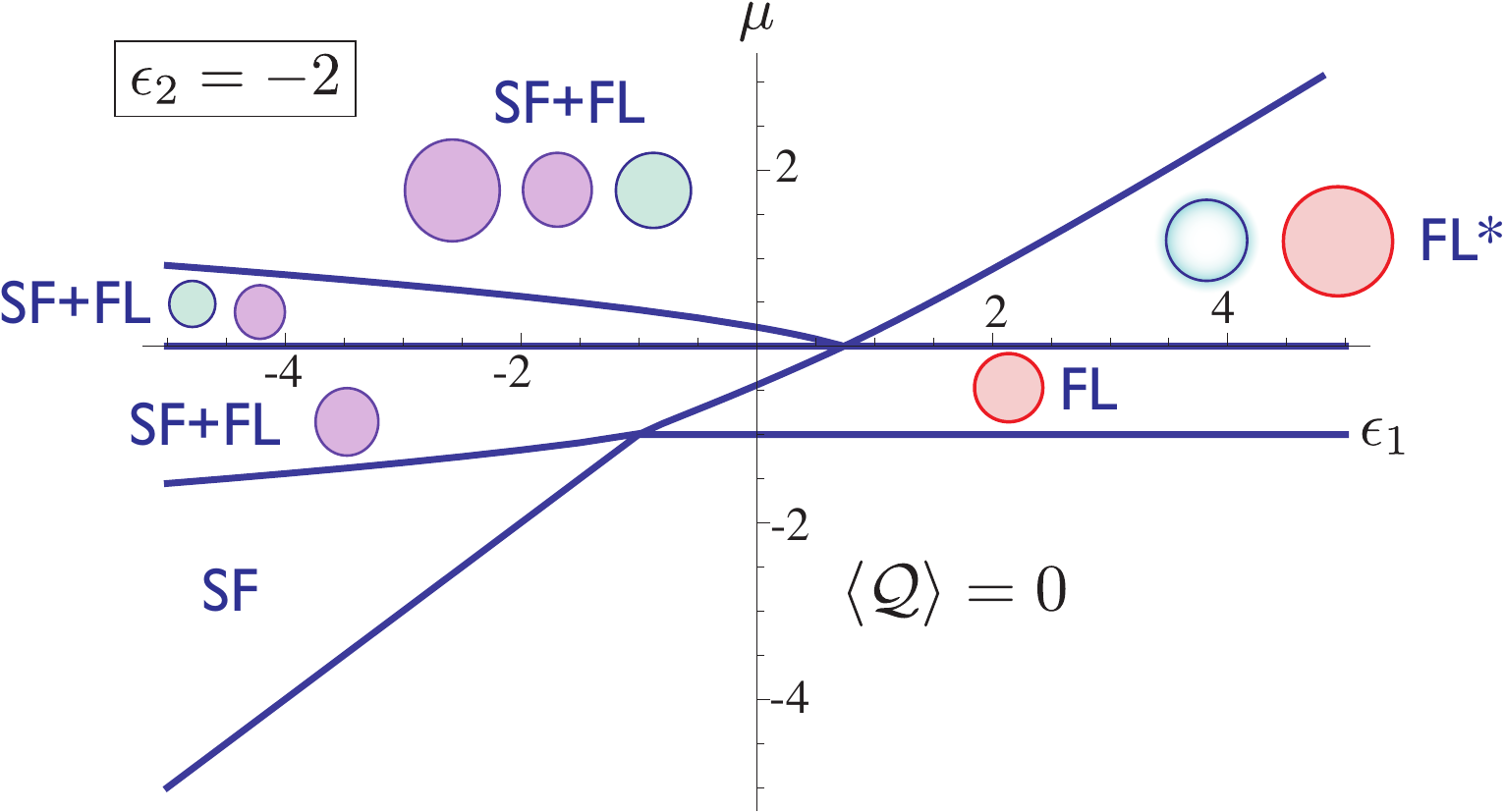}
\end{center}
\caption{As in Fig.~\ref{pd_sym3}, but with a different value of $\epsilon_2$.}
\label{pd_sym4}
\end{figure}
For
$\mu_1 = \mu_2 \neq 0$, $\mu_3 =0$, the charged scalars and fermions both have unit global U(1) charges. This is in contrast
to the case already considered, where the scalars were doubly charged. Consequently it is no longer possible to 
mix the scalar with fermion pairs. There also remain some fermions with a relativistic spectrum which are neutral
under the U(1) charge, but we neglect these in the spirit of the simplifications we have used. Following the same reasoning as above,
we are then led to the following theory for this chemical potential assignment
\bea \widetilde{\mathcal{L}}_{3} &=&  \widetilde{\mathcal{L}}_{\Phi} + \widetilde{\mathcal{L}}_\lambda
+ \widetilde{\mathcal{L}}_c    \nn
\widetilde{\mathcal{L}}_\Phi &=& \Phi^{\dagger} \left( D_{\tau} - \mu + \epsilon_{1}
 -  \frac{{\bm D}^{2}}{2m_1} \right)  \Phi + 
u  \left( \Phi^{a\dagger} \Phi^{a} \right)^2 \nn
\widetilde{\mathcal{L}}_\lambda &=&  \lambda^{\dagger} \left(D_{\tau} - \mu -  \frac{{\bm D}^{2}}{2m_2}  \right)  \lambda \nn
\widetilde{\mathcal{L}}_c &=&  c^{\dagger} \left(\partial_{\tau} - 2 \mu + \epsilon_{2} -  \frac{{\bm \nabla}^{2}}{2m_3}    \right) c 
+  g_2 \left(  c^{\dagger} \lambda^a \Phi^{a} + {\rm c.c.} \right). \label{tildeL}
\eea
Note the absence of the pairing $g_1$ term which was present in Eq.~(\ref{L0}); this is because it is prohibited by 
the conserved global U(1) charge which is modified from Eq.~(\ref{qsym}) to 
\beq
\mathcal{Q} = \lambda^{a\dagger} \lambda^{a} + \Phi^{a \dagger} \Phi^{a} + 2 c^{\dagger} c. \label{qsym2}
\eeq
The mean-field phases of this theory can be obtained as above: we only have to modify the mean field equations by setting $g_1=0$
and use the different chemical potential assignments. Indeed, the analyses and phases turn out to be very similar to
the ABJM-inspired models in Section~\ref{sec:ABJM} with $g_1=0$ in Eq.~(\ref{l0a}). The resulting phase diagrams in
Figs.~\ref{pd_sym3} and~\ref{pd_sym4} are very similar to those in Figs.~\ref{pd_sym1} and~\ref{pd_sym2}.

The non-SF phases, which have $\langle \Phi^a \rangle = 0$, have Fermi surfaces which obey a Luttinger constraint descending
from Eq.~(\ref{qsym2}):
\beq
3 V_\lambda + 2 V_c = \langle \mathcal{Q} \rangle;
\eeq
Here $V_\lambda$ is the volume enclosed by the $\lambda^a$ Fermi surface, and the prefactor 3 arises
from the summation over the $a$ index. The new feature of Figs.~\ref{pd_sym3} and~\ref{pd_sym4}
is the presence of deconfined phases with Fermi surfaces of $\lambda^a$ fermions
coupled to gapless gauge field, the FL* and NFL phases; such phases were absent with the different chemical potential assignments
in Figs.~\ref{pd_sym1} and~\ref{pd_sym2}. As in Sections~\ref{sec:doublon} and~\ref{sec:ABJM}, the NFL and FL* phases are 
expected to be eventually unstable to fermion pairing induced by gauge boson fluctuation \cite{sonqcd}, and it would interesting to study the strength of this instability in the large $N$ limit of the gauge theory.

\section{Discussion}
\label{sec:dis}

This paper has drawn connections between the compressible quantum states of models studied in condensed matter,
to those in models amenable to studies by gauge-gravity duality. 

In Sections~\ref{sec:doublon},~\ref{sec:bf}, and~\ref{sec:ffl} we presented a unified
discussion of previously studied condensed matter models, which contain a full range of compressible phases. As discussed in Section~\ref{sec:intro}, compressible phases which do not break a global U(1) symmetry associated with a charge $\mathcal{Q}$
must have Fermi surfaces whose total volume is constrained by the value of $\langle \mathcal{Q} \rangle$. The most common compressible phase is, of course, the familiar Fermi liquid (FL). However, we also found non-Fermi liquid (NFL) phases in which
the Fermi surface quasiparticles were coupled to Abelian or non-Abelian gauge fields; in both cases, the damping of the gauge modes by Fermi surface excitations is expected to stabilize a deconfined phase of the gauge theory.  Finally, we found fractionalized Fermi liquid phase (FL*), which may be viewed as a co-existence of FL and NFL phases, with Fermi surfaces of both gauge neutral
and gauge charged particles. The FL* phase is crucial for studies of gauge-gravity duality, because it provides a route for strong scattering of gauge-neutral particles: the Green's functions of such gauge-neutral particles appear as observables in the dual gravity theory. Schematic phase diagrams of such phases in the condensed matter models appear in 
Figs.~\ref{pd_nfl},~\ref{pd_bf}, and~\ref{pd_ffl}.

Next, in Sections~\ref{sec:ABJM} and~\ref{sec:SYM} we examined two of the workhorses of gauge-gravity duality:
the ABJM model in spatial dimension $d=2$, and the SYM theory in $d=3$. In both cases, we perturbed the conformal field
theory with a chemical potential, and used the structure of the resulting theory to motivate toy models of compressible quantum matter. We presented the phase diagrams of these models in Figs.~\ref{pd_abjm0}--\ref{pd_sym4}. While the detailed
patterns of gauge and global charges were somewhat different from the condensed matter models, the basic phases were the same.
In particular, the compressible phases with $\mathcal{Q}$ symmetry preserved were FL, NFL, and FL*.

A related correspondence between the condensed matter models and the string-inspired models was made in Ref.~\onlinecite{ssffl}.
This correspondence begins with the lattice
discretization of the continuum theory $\mathcal{L}_\ast$ in Eq.~(\ref{Ls}), which is reviewed in Ref.~\onlinecite{tasi}. 
The resulting lattice model is solvable in the limit of infinite-range hopping, or infinite $d$, combined with a particular large $N$ 
limit \cite{sy,pgs,bgg,si,sigeorge}.
It was shown that the physical properties of this solvable model coincided with those of the classical dual gravity model of
Ref.~\onlinecite{hong1}; in the latter model the classical gravity approximation led to a theory on the space AdS$_{2} \times$R$^d$. 
Specifically \cite{statphys}, both models had compressible phases with non-zero ground state entropy density, correlations which had momentum-independent singular temporal correlations with the structure of conformal quantum mechanics, and singular damping of the gauge-neutral particles at the $c$ Fermi surface. It was proposed \cite{ssffl}, therefore, that the gravity theory of
Ref.~\onlinecite{hong1} had realized an infinite-range limit of the FL* phase.

However, most of the physical properties of the FL* phase so obtained are expected to be consequences of the respective 
simplifications: the infinite-range limit in the condensed matter models, and the factorized AdS$_{2} \times$R$^d$ geometry
in the classical gravity theory. Nevertheless, it is quite remarkable that two very different solvable limits lead to essentially the same physical properties,
which could apply to physical systems over a significant intermediate energy scale \cite{ssffl}.

The challenge for the future is to describe the phases of Sections~\ref{sec:ABJM} and~\ref{sec:SYM} using the dual gravity theory,
in a manner which captures their expected properties of finite-range models in $d=2$ and $d=3$ respectively, and there have been recent
studies in this direction \cite{gubserrocha,kiritsis,sean1,sean2,sean3,eric,kachru2,kachru3,trivedi}. 
In particular, the $d=3$ model of Gubser and Rocha \cite{gubserrocha} is a promising model for future study. 
The problem of a Fermi surface coupled to a gauge field
appears to be under control in $d=3$: the results of the self-consistent one-loop theory \cite{pincus,reizer1,reizer2} are expected to be robust to  higher order corrections \cite{leen,metnem}. Such a theory only gives marginal corrections to the FL results: the low $T$
specific heat behaves as $T \log T$. It would be interesting to see if such corrections eventually emerge from dual gravity theories of compressible
matter in $d=3$, such as that of Ref.~\onlinecite{gubserrocha}. 

\acknowledgements
We are grateful to Max Metlitski for a number of key comments at the initial stages of this work.
We thank E.~Berg, S. Gubser, S. Hartnoll, D.~Hofman, S.~Kachru, I. Klebanov, H.~Liu, J. Maldacena, C.~Mathy, K.~Rajagopal, E.~Silverstein, D.~Son, S.~Trivedi, and E. Witten for valuable discussions.
This research was supported by the National Science Foundation under grant DMR-0757145, by a MURI grant from AFOSR, and by the Netherlands Organisation for Scientific Research (NWO).

\appendix

\section{Lagrangian for $\mathcal{N}=4$ Super Yang-Mills theory}
\label{app:SYM}

This appendix will write down the Lagrangian for the Yang-Mills theory in $d=3$ spatial
dimensions with $\mathcal{N}=4$ supersymmetry. We will use a non-relativistic notation, with all indices explicit,
designed to address the case with non-zero chemical potential. Thus relativistic invariance, supersymmetry, and the associated global SU(4) symmetry
will not be explicit: this is acceptable because the chemical potential breaks these symmetries anyway.

First, let us note the particle content of the theory with a SU(N) gauge group.
\begin{itemize}
\item The fermions are complex 2-component Weyl spinors $\lambda_{i\alpha}^a$,
with the
Weyl index $\alpha = 1,2$. The fermions transform as the adjoint of the gauge group SU($N$), and the
color index $a = 1 \ldots N^{2}-1$. Without any chemical potentials, there is a SU($4$) global symmetry, and the fermions
transform as the fundamental of SU($4$) with $i = 1 \ldots 4$.
\item There complex scalars $\Phi_p^a$ are also adjoints of color SU($N$) with $a =1 \ldots N^{2}-1$.
They transform as a real 6-dimensional representation of SU(4), and so the index $p=1,2,3$.
\item The gauge field $(A_{\tau}^{a}, {\bf A}^{a})$ has a color index $a$.
\end{itemize}

It is possible to add 3 chemical potentials coupling to commuting generators of the global SU($N$). We choose\cite{yamada} these chemical
potentials, $\mu_{p}$, so that their couplings to the scalar field are diagonal in the $p$ index. Then, the imaginary time kinetic
term for the scalar field is
\bea 
\mathcal{L}_\Phi &=& \sum_{p=1}^{3} \biggl\{
\Bigl( \left[ (\partial_\tau + \mu_{p})\delta_{ab} - f_{acb}A_{\tau}^{c}\right] \Phi_p^{b \dagger} \Bigr)
\Bigl( \left[ (\partial_\tau - \mu_{p})\delta_{ad} - f_{aed}A_{\tau}^{e}\right] \Phi_p^{d} \Bigr)   \nn
&~& \quad \quad \quad + \,  \Bigr(\left[ {\bm \nabla} \delta_{ab} - f_{acb} {\bf A}^{c}\right] \Phi_p^{b\dagger}\Bigr) \Bigr(\left[ {\bm \nabla} \delta_{ad} - f_{aed} {\bf A}^{e}\right] \Phi_p^{d}\Bigr) 
\biggr\}.
\eea

Symmetry now dictates how the chemical potentials couple to the fermions.\cite{yamada} The fermions kinetic energy terms in imaginary time are
\bea
\mathcal{L}_\lambda &=& \sum_{i=1}^{4} \lambda_{i\alpha}^{a \dagger} \Bigl( (\partial_\tau - \widetilde{\mu}_{i} ) \delta_{ab}
 - f_{acb}A_{\tau}^{c}
 + i {\bm \sigma}_{\alpha\beta} \cdot \left( {\bm \nabla} \delta_{ab} - f_{acb} {\bm A}^{c} \right) 
 \Bigr) \lambda_{i\beta}^b 
\eea
Here ${\bm \sigma}$ are the Pauli matrices, and the fermion chemical potentials are
\bea \widetilde{\mu}_{1} &=& \frac{\mu_{1}+\mu_{2}+\mu_{3}}{2} \nn
 \widetilde{\mu}_{2} &=& \frac{\mu_{1}-\mu_{2}-\mu_{3}}{2} \nn
 \widetilde{\mu}_{3} &=& \frac{-\mu_{1}+\mu_{2}-\mu_{3}}{2} \nn
 \widetilde{\mu}_{4} &=& \frac{-\mu_{1}-\mu_{2}+\mu_{3}}{2} .
\eea

There is a standard Yang-Mills kinetic term, $\mathcal{L}_{A}$, for the SU($N$) gauge field and we will not display this explicitly.

Most crucial for our purposes are the Yukawa couplings between the scalars and the fermions. We write these as
\bea
\mathcal{L}_Y &=& g f_{abc} \varepsilon_{\alpha\beta} \Bigl( \Phi_1^{a \dagger} \lambda_{1\alpha}^b \lambda_{2\beta}^c  + \Phi_1^{a} \lambda_{3\alpha}^b \lambda_{4\beta}^c \nn
&~& \quad \quad \quad + \, \Phi_2^{a \dagger} \lambda_{1\alpha}^b \lambda_{3\beta}^c  + \Phi_2^{a} \lambda_{4\alpha}^b \lambda_{2\beta}^c \nn
&~&\quad \quad \quad + \, \Phi_3^{a \dagger} \lambda_{1\alpha}^b \lambda_{4\beta}^c  + \Phi_3^{a} \lambda_{2\alpha}^b \lambda_{3\beta}^c + \mbox{c.c.} \Bigr)
\label{LY}
\eea
Here $g$ is the single coupling constant of the theory. It is easy to check that these couplings are invariant under the `diagonal' SU(4) transformations
and the associated chemical potential assignments to the scalars and fermions. Less explicit is the symmetry of the Yukawa couplings under
the off-diagonal SU(4) transformations. It can be checked that $\mathcal{L}_{Y}$ is invariant under the following transformations
\bea 
\delta \lambda_1 &=& i \lambda_2 \quad , \quad \delta \lambda_2 = i \lambda_1 \quad , \quad\delta \Phi_3 = -i \Phi_2^\dagger \quad , \quad \delta \Phi_2  =  i \Phi_3^\dagger \nn 
\delta \lambda_1 &=&   \lambda_2 \quad , \quad \delta \lambda_2 = - \lambda_1 \quad , \quad\delta \Phi_3 = -  \Phi_2^\dagger \quad , \quad \delta \Phi_2  =   \Phi_3^\dagger \nn 
\delta \lambda_1 &=& i \lambda_3 \quad , \quad \delta \lambda_3 = i \lambda_1 \quad , \quad \delta \Phi_1  =   -i \Phi_3^\dagger 
 \quad , \quad\delta \Phi_3 =  i \Phi_1^\dagger\nn 
\delta \lambda_1 &=&  \lambda_3 \quad , \quad \delta \lambda_3 = - \lambda_1  \quad , \quad \delta \Phi_1  =  - \Phi_3^\dagger 
\quad , \quad\delta \Phi_3 =   \Phi_1^\dagger\nn 
\delta \lambda_1 &=& i \lambda_4 \quad , \quad \delta \lambda_4 = i \lambda_1 \quad , \quad\delta \Phi_2 = - i \Phi_1^\dagger \quad , \quad \delta \Phi_1  =   i \Phi_2^\dagger \nn 
\delta \lambda_1 &=&  \lambda_4 \quad , \quad \delta \lambda_4 = - \lambda_1 \quad , \quad\delta \Phi_2 =   -\Phi_1^\dagger \quad , \quad \delta \Phi_1  =   \Phi_2^\dagger \nn 
\delta \lambda_2 &=& i \lambda_3 \quad , \quad \delta \lambda_3 = i \lambda_2 \quad , \quad\delta \Phi_2 =  i \Phi_1 \quad , \quad \delta \Phi_1  =   i \Phi_2 \nn 
\delta \lambda_2 &=&  \lambda_3 \quad , \quad \delta \lambda_3 = - \lambda_2 \quad , \quad\delta \Phi_2 =   -\Phi_1 \quad , \quad \delta \Phi_1  =   \Phi_2 \nn 
\delta \lambda_3 &=& i \lambda_4 \quad , \quad \delta \lambda_4 = i \lambda_3 \quad , \quad\delta \Phi_3 = i \Phi_2 \quad , \quad \delta \Phi_2  =  i \Phi_3 \nn 
\delta \lambda_3 &=&   \lambda_4 \quad , \quad \delta \lambda_4 = -  \lambda_3 \quad , \quad\delta \Phi_3 =  - \Phi_2
\quad , \quad \delta \Phi_2  =   \Phi_3 \nn 
\delta \lambda_4 &=& i \lambda_2 \quad , \quad\delta \lambda_2 = i \lambda_4  \quad , \quad \delta \Phi_1  =  i \Phi_3 
\quad , \quad \delta \Phi_3 = i \Phi_1\nn 
\delta \lambda_4 &=&   \lambda_2 \quad , \quad \delta \lambda_2 = -  \lambda_4 \quad , \quad\delta \Phi_1 = - \Phi_3 \quad , \quad \delta \Phi_3  =  \Phi_1 ,
\eea
which combine to yield full SU(4) symmetry. Note that the chemical potential terms in $\mathcal{L}_{\Phi}$ and $\mathcal{L}_{\lambda}$
are not invariant these off-diagonal transformations.

There are also quartic interactions between the scalars which we will not write out, because they are not important for our purposes.
This is because the chemical potentials modify the scalar potentials, and so the special restrictions of supersymmetry on the form
of the scalar potential has no bearing on our considerations.

With all the terms in the action described, we are now ready to discuss the structure of the fermion pairing terms in $\mathcal{L}_{Y}$.
It is useful to discuss these in using canonical fermion operators near the Fermi level.
We therefore make the following mode expansion \cite{krishna1} in terms of canonical Fermi 
operators $a_{i}^{a} ({\bf k})$ and $b_{i}^{a} ({\bf k})$
\beq
\lambda_{i}^{a} ({\bf x}) = \int \frac{d^{3}k}{(2 \pi)^{3}} \left( \begin{array}{c} - \sin( \theta_{\bf k}/2) e^{-i \phi_{\bf k}} \\
\cos( \theta_{\bf k}/2) \end{array} \right) \left[ a_{i}^{a} ({\bf k}) e^{- i {\bf k} \cdot {\bf x}} + b_{i}^{a \dagger} ({\bf k}) e^{ i {\bf k} \cdot {\bf x}} \right], \label{mode1}
\eeq
where $\theta_{\bf k}$ and $\phi_{{\bf k}}$ are the polar and azimuthal angles of ${\bf k}$, so under ${\bf k} \rightarrow - {\bf k}$, $\theta \rightarrow \pi - \theta$ and $\phi \rightarrow \phi + \pi$.
The single-particle Hamiltonian for these canonical Fermi fields is
\beq
H_{\lambda} = \sum_{i=1}^{4} 
\int \frac{d^{3}k}{(2 \pi)^{3}} \biggl[ \Bigl(|{\bf k}| - \widetilde{\mu}_{i} \Bigr) a_{i}^{a \dagger} ({\bf k}) a_{i}^{a} ({\bf k}) + \Bigl( |{\bf k}| + \widetilde{\mu}_{i} \Bigr) b_{i}^{a \dagger} ({\bf k}) b_{i}^{a} ({\bf k}) 
\biggr]
\eeq 
The fermion pair terms in $\mathcal{L}_{Y}$ are $f_{abc} \varepsilon_{\alpha\beta} \lambda_{i\alpha}^{b} \lambda_{j \beta}^{c}$, and these
are antisymmetric in color, SU(4) `flavor', and Weyl spin. The dominant pairing will arise from fermions at the same Fermi energy.
We will mainly consider the case where only one chemical potention, say $\mu_{1}>0$, is non-zero. 
Then the Fermi level excitations are $a_{1}^{a}$, $a_{2}^{a}$, 
and $b_{3}^{a}$ and $b_{4}^{a}$, and we ignore the remaining fermions. Then we have
\beq
\varepsilon_{\alpha\beta} \left\langle  \lambda_{1\alpha}^{b} \lambda_{2 \beta}^{c} \right\rangle = 
\int \frac{d^{3}k}{(2 \pi)^{3}} e^{-i \phi_{\bf k}} \left \langle a_{1}^{a} (-{\bf k}) a_{2}^{b} ({\bf k}) \right\rangle
\eeq
So we should have\cite{krishna1,krishna2}
\beq
\left \langle a_{1}^{a} (-{\bf k}) a_{2}^{b} ({\bf k}) \right\rangle \propto e^{i \phi_{\bf k}} f_{abc}
\eeq
to obtain pairing that is antisymmetric in color, SU(4) flavor, and Weyl spin. The $e^{i \phi_{\bf k}}$ factor above represents the antisymmetry in Weyl spin for our choices for the fermion normal modes in \eqnref{mode1}.
Such pairing which is antisymmetric in color, flavor, and spin is just that expected from the attractive interaction from the
SU($N$) gauge force.\cite{krishna1,krishna2}

\section{Phases with $\langle b_+ \rangle \neq 0$ and $\langle b_- \rangle = 0$.}
\label{app:gauge}

\begin{figure}[h]
\begin{center}
\includegraphics[width=6in]{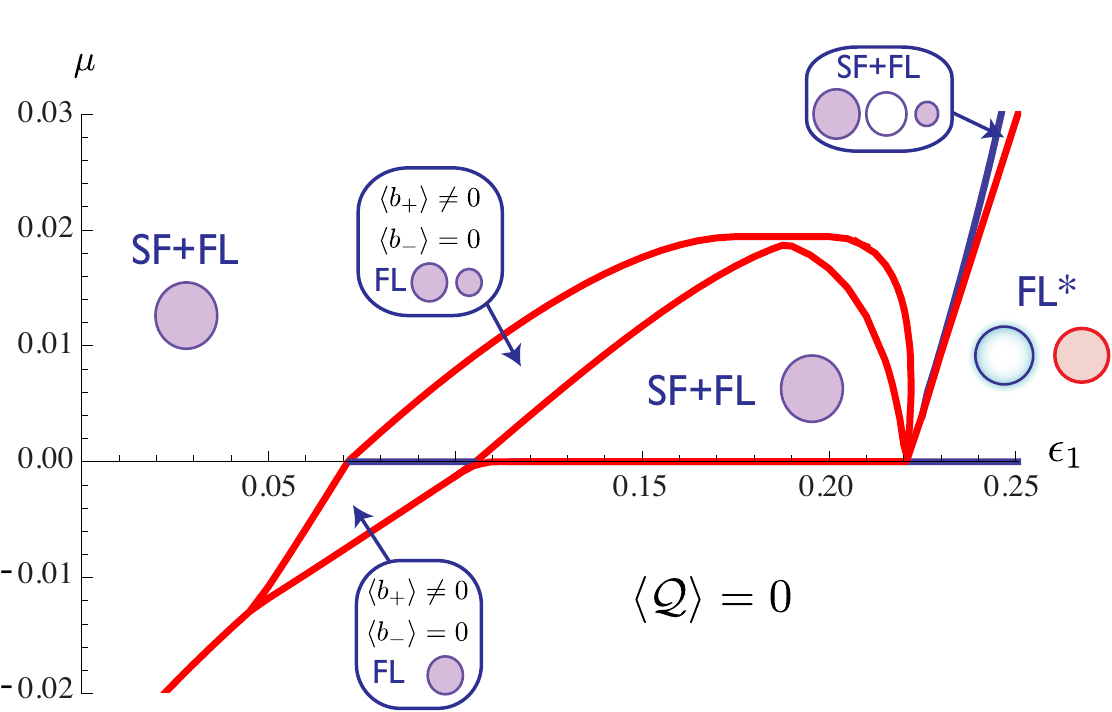}
\end{center}
\caption{Mean-field phase diagram of $\mathcal{L}_1$ as in Fig.~\ref{pd_abjm5}, with $g_1 = 1$, $g_2 = 1$, but with $v=4$ and $\epsilon_2=0$. We zoomed in on a region of the phase diagram where the FL phases with $\langle b_+ \rangle \neq 0$ and $\langle b_- \rangle = 0$ appear.
The SF+FL phase with 3 Fermi surfaces is similar to that in Fig.~\ref{pd_abjm5}.}
\label{pd_abjm7}
\end{figure}

In the mean-field analysis of the theory analogous to the ABJM model $\mathcal{L}_1$ in Eq. (\ref{l1}) the confining phase with $\langle b_+ \rangle \neq 0$ or $\langle b_- \rangle \neq 0$, but not both, appears only for small values of the chemical potential $\mu$ and the detuning $\epsilon_2$. In this appendix we discuss this phase in a bit more detail.

In the following we will assume that in the confining phase $b_+$ condenses and $b_-$ does not. The gauge neutrality constraint (\ref{bfconst}) then reads
\beq
\langle Q_c \rangle =| b_+|^2  + \langle f_+^\dagger f_+ \rangle - \langle f_-^\dagger f_- \rangle=0.
\eeq

From (\ref{f1f2}) we find $\langle f_+^{\dag} f_+ \rangle=\langle f_2^{\dag} f_2 \rangle$ and $\langle f_-^{\dag} f_- \rangle=\langle f_1^{\dag} f_1 \rangle$ in the phase where $\langle b_- \rangle = 0$. In this phase the $f_2$ fermion decouples so we have
\beq
\langle f_2^{\dag} f_2 \rangle = \int \frac{d^2 k}{(2 \pi)^2} \theta (-\xi_k^f),
\eeq
where $\xi_k^f= k^2/(2 m_f)-\mu$. Consequently, $\langle f_2^{\dag} f_2 \rangle=m_f \mu/ (2 \pi) \theta( \mu)$. The $f_1$ fermion hybridizes with the $c$ fermion. We define the hybridized fermions $F_{\pm}$ through their dispersion relation:
\bea
 \xi^{F_{\pm}}_k &=& \frac{1}{2} (\xi^c_k + \xi^{f}_k) \pm \frac{1}{2} \sqrt{ (\xi^c_k - \xi^{f}_k)^2 + 4 g_2^2 |b_+|^2} ,
\eea
where $\xi_k^c= k^2/(2 m_c)-2\mu +\epsilon_2$. For the $f_1$ fermions we then have
\beq
\langle f_1^{\dag} f_1 \rangle = \int \frac{d^2 k}{(2 \pi)^2} ( \cos^2 (\theta_k) \theta (-\xi_k^{F_+}) + \sin^2 (\theta_k) \theta (-\xi_k^{F_-}) ),
\eeq
where $\theta_k$ is the mixing angle in the unitary transformation:
\beq
\left( \begin{array}{c}  F_- \\ F_+ \end{array} \right) = \left( \begin{array}{cc}  \cos(\theta) & \sin(\theta) \\ \sin(\theta) & - \cos (\theta) \end{array} \right) \left( \begin{array}{c} c \\ f_1 \end{array} \right) .
\eeq
We find
\bea
\sin^2 (\theta_k) &=& \frac{1}{2}+\frac{(\xi^c_k - \xi^{f}_k)}{2 \sqrt{ (\xi^c_k - \xi^{f}_k)^2 + 4 g_2^2 |b_+|^2}} \nn
\cos^2 (\theta_k) &=&\frac{g_2^2 |b_+|^2}{| (\xi^c_k - \xi^{f}_k)^2 + 4 g_2^2 |b_+|^2| \sin^2 (\theta_k) } .
\eea
With this we can compute the total gauge charge $\langle Q_c \rangle$ as a function of $|b_+|$ and impose the condition that it is zero.

\begin{figure}
\centering
\includegraphics[width=0.8\textwidth]{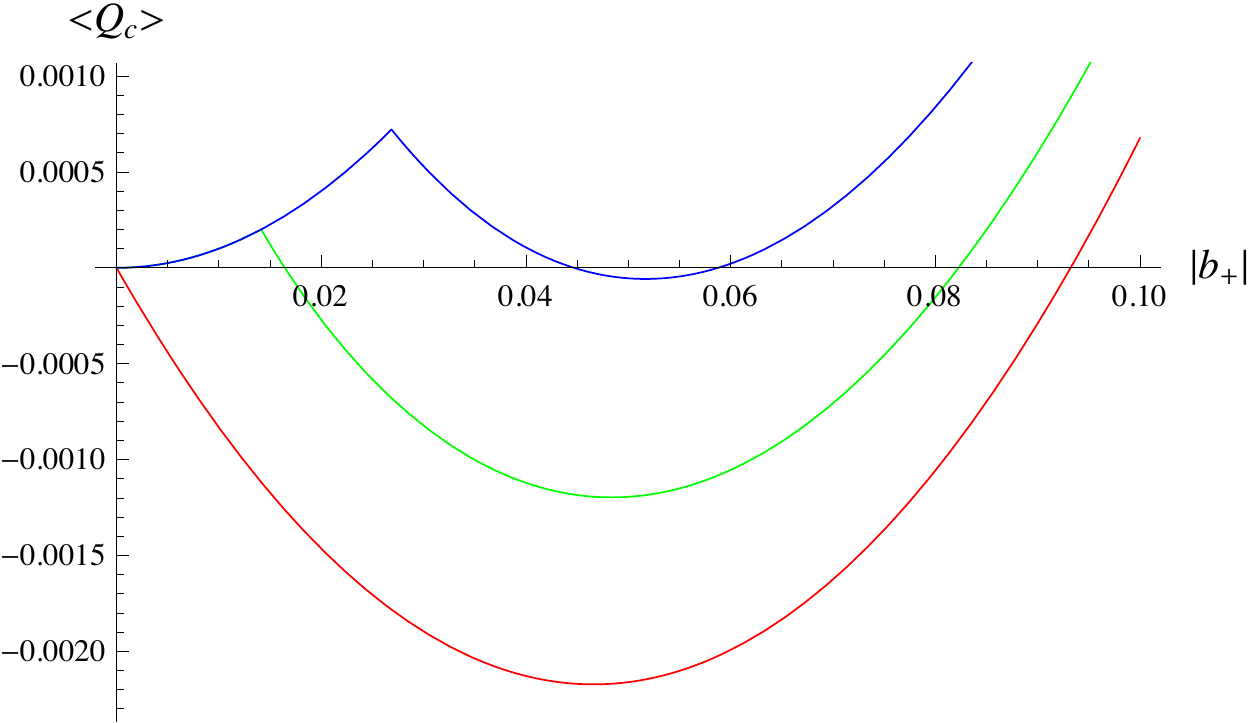}
\caption{We plot $\langle Q_c \rangle$ as a function of $|b_+|$ for $\epsilon_2=0$ and $\mu=0$ (red line), $\mu=\pm0.01$ (green line) and $\mu=\pm0.019$ (blue line).\label{fig:gc}}
\end{figure}

Clearly, $|b_+|=0$ always satisfies the constraint. We find, however, that the condition is only satisfied for $|b_+|>0$ when both $\epsilon_2$ and $\mu$ are close to zero. As an example, for $\epsilon_2=0$ there are solutions for $|b_+|>0$ when $|\mu| \lesssim 0.02$ (see Fig. \ref{fig:gc}).

In figure \ref{pd_abjm7} we show the phase diagram as a function of $\epsilon_1$ and $\mu$ where $\epsilon_2=0$. We zoomed in on the region of small $\mu$. For $\mu>0$ the region with $\langle b_+ \rangle \neq 0$ and $\langle b_- \rangle = 0$ has a funny curved shape, which is explained by the fact that for $\mu<0.0193$ there are two solutions to the gauge constraint with $|b_+|>0$. Although this is also true for $\mu<0$, we find that in that region the minimum always occurs at the larger solution of $|b_+|$. This is related to the fact that for both detunings and the chemical potential close to zero the condensation into this phase without a gauge constraint becomes first order.\cite{marchetti}

\end{document}